\documentclass[twocolumn]{aastex631}

\usepackage{url}
\usepackage{graphicx}
\usepackage{natbib}
\usepackage{textcomp}
\usepackage{gensymb}
\usepackage{enumitem}
\usepackage{xcolor}
\usepackage{hyperref}

\begin{document}

\title{The Many-Faceted Light Curves of Young Disk-bearing Stars in Taurus as Seen by {\em K2}}

\correspondingauthor{Ann Marie Cody} 
\email{acody@seti.org} 

\author{Ann Marie Cody}
\affiliation{SETI Institute, 339 N Bernardo Ave, Suite 200, Mountain View, CA 94043}

\author{Lynne A. Hillenbrand}
\affiliation{Department of Astronomy, California Institute of Technology, Pasadena, CA 91125, USA}

\author{Luisa M. Rebull}
\affiliation{Infrared Science Archive (IRSA), IPAC, 1200 E. California Blvd., California Institute of Technology, Pasadena, CA 91125, USA}

\begin{abstract}

We present a comprehensive study of the variability properties of young disk-bearing stars in the Taurus star-forming region, paralleling our previous \citep{cody2018} investigation in $\rho$ Oph and Upper Sco.  A sample of 99 confirmed Taurus association members is placed in the diagnostic $Q-M$ plane of flux asymmetry ($M$) and quasi-periodicity ($Q$), which guides our assignment of variability classes. We find a similar proportion of flux-symmetric variables in Taurus, but more bursters and fewer dippers relative to Upper Sco. The regions also differ in that the amplitudes for periodic and quasi-periodic sources are larger in Taurus relative to the more evolved Upper Sco star/disk systems.  The relationship between photometric variability patterns at optical wavelengths, which arise in the inner disk and at the stellar surface, are assessed relative to available disk inclination measurements.

\end{abstract}


\section{Introduction}

Young stars are photometric variables  \citep{joy1945,herbig1954,walker1954}
on timescales that range from hours to weeks, with a typical timescale of 1-2 days \citep{costigan2014,findeisen2015t}.
The variability is observed over a broad range of wavelengths \citep[e.g.][]{venuti2015,herbst2002,carpenter2001,rebull2014}.
In the optical range, the dominant effects associated with the star are due to rotational modulation of
active ``starspot" regions in the photosphere, as well as magnetically induced chromospheric activity 
at higher altitudes in the atmosphere.
However, for those young stars still surrounded by circumstellar material, there is also
variability due to  unsteady accretion from the inner circumstellar disk onto the star (causing brightening)
and, in some cases, due to changes in line-of-sight extinction (causing dimming). 

Modelling efforts aimed at understanding the morphologies observed in high-cadence light curves of young disk-bearing stars include  those by \cite{kesseli2016} and \cite{robinson2021}, among others.
Much of the accretion-related and extinction-related variability originates in the star-disk interaction region.
This zone encompasses the accretion shock on the surface of the star, the magnetosphere in which accreting gas and dust are entrained,  and the innermost regions of the highly irradiated circumstellar disk.  

Empirically, the diversity of light curve shapes for young stars, and their characteristic timescales, was appreciated only with the availability of high-cadence and very high-precision photometric observations from spacecraft, specifically {\em CoRoT} \citep{baglin2006}, {\em MOST} \citep{matthews2000}, and {\em K2} \citep{howell2014}. 
\cite{cody2014} defined a classification scheme to characterize young variables using a scale from strictly periodic, to quasi-periodic, to entirely stochastic, on one axis, and on the other, quantifying primarily bursting, symmetric, and primarily fading/dipping behavior.  
These dubbed $Q$ and $M$ metrics were developed assuming high quality photometric times series data.

{\em CoRoT} provided the first such high-cadence, high-precision studies of young stars. The NGC 2264 region was observed over 22 days in 2008 and 39 days in 2011, sampling $\sim$175 young cluster members \citep{cody2014,stauffer2014,stauffer2015,mcginnis2015,stauffer2016,sousa2016,venuti2017}. The next opportunity came with coverage of the Upper Scorpius and $\rho$~Ophiuchus regions in {\em K2} Campaigns 2 and 15.
These observations provided light curves over $\sim$80 days for $\sim$1300 (C2) and $\sim400$ (C15) young cluster members, respectively
\citep{rebull2018,cody2017,cody2018,ansdell2018}. {\em K2} data is also available for the Lagoon Nebula Cluster, although mainly for higher mass stars \citep{venuti2021}. 

We present here the fourth young region with space-based optical light curves. Our study uses {\em K2} Campaign 4 and 13 coverage of the northern portion of the Taurus region, with $\sim$70 and $\sim$80 days of photometric time series data, respectively.  A total of 181 young cluster members were targeted (not all of which have high-quality data). 

We note that the Transiting Exoplanet Survey Satellite \citep[TESS;][]{ricker2015} is also now contributing high-cadence, high-precision light curves of young stars. Its all-sky coverage encompassing many young clusters including the extended Taurus regions, albeit to brighter limiting magnitudes than the {\em K2} or {\em CoRoT} data sets described above.

The greater Taurus region has been the subject of many previous photometric monitoring studies over the past eight decades, conducted using ground-based telescopes. There is a rich literature concerning the variability properties of individual stars on timescales from days to decades. The entire history of such studies is too vast to review here,  but includes long-term monitoring programs e.g. by \cite{herbst1994} and \cite{grankin2007}. Wide-field imaging survey results were first provided by \cite{slesnick2006} based on Palomar/QUEST data, and \cite{xiao2012} based on TrES data, followed by \cite{rodriguez2017} who used KELT data.
The opportunity presented by the {\em K2} mission, however, is the first possible investigation of the young Taurus population with precision photometry.
Only a few of the Taurus members have benefited previously from such high-quality data, via observations with {\em MOST} \citep{cody2013}.

From the same set of {\em K2} data in Taurus that we are using, \cite[][hereafter ``R20'']{rebull2020} presented the sample of periodic sources, and analyzed the period-color diagram in the context of stellar rotational evolution during the pre-main sequence. In addition, \citet{roggero2021} have analyzed in detail the subset of {\em K2} Taurus ``dipper" stars-- those that display fading events in their light curves characteristic of dust occultation. {\em K2} photometry of Taurus members has also appeared in several other papers on individual objects.
\cite{vandam2020} studied the evidence for an eclipsing substellar companion to the binary V928 Tau, a weak-line T Tauri system. \cite{pouilly2020} studied the early type accreting star HQ Tau and \citet{pouilly2021} the late-type dipper variable V807~Tau. \cite{biddle2021} studied the classical T Tauri star CI Tau, \cite{kospal2018} investigated the accreting young binary DQ~Tau, and \cite{2018ApJ...861...76P} examined flares associated with brown dwarf CFHT-BD-Tau 4.

In this paper, we focus on the disk-bearing sample of Taurus. We perform an analysis similar to that in \cite{cody2018},
classifying the light curve behaviors into morphological groups that are related to the influences of rotation, accretion, and circumstellar dust on the time-dependent photometric brightness of the sources. We also compare our results for Taurus to our previous findings for disk-bearing samples of young stellar objects in NGC 2264, $\rho$~Oph, and Upper Sco.

\section{The sample of young disk-bearing stars in Taurus}

The {\em Kepler} Space Telescope observed Taurus as part of its 4th and 13th {\em K2} Campaigns from 2015 February 7 to 2015 April 23, and 
2017 March 8 to May 27, respectively. While many known Taurus members from the literature were deliberately included in the Campaigns, a handful of previously unknown young stars also fell within the {\em K2} field of view. 

We began with a list of 216 known and candidate Taurus members compiled as part of our Campaign 13 target request that had been submitted to the mission. Once the approved targets from all observing programs were released, we recognized that there may be additional young stars serendipitously included as {\em K2} targets by other groups, e.g., as planet search targets. Consequently, R20 used the Gaia mission second data release (Gaia DR2) to mine the complete set of all stars observed by {\em K2} in Campaigns 4 and 13 for colors, magnitudes, parallaxes, and proper motions consistent with Taurus membership. They ultimately published a set of 156 Taurus members observed by {\em K2}, along with 23 additional candidates of less certain status. This same sample serves as the basis set for the current study.

In this work, we are interested in assessing the variability of stars whose inner circumstellar disks have not yet dissipated. Here, we have adopted the disk selection criteria of R20, who used spectral energy distributions to identify objects in the member and candidate member samples with infrared excesses. Based on their assessment, we find a total of 99 disk-bearing stars in the {\em K2} Taurus sample, of which 93 have definitive inner disks, and the remaining six have smaller excesses which may also indicate circumstellar dust. Stars in this latter group are included as disk-bearing ``candidates" and are noted as such in Table~1.

Relative to previous samples with high-cadence, high-precision light curve data sets, the $K2$ Taurus sample is slightly smaller. $\rho$~Oph had 123 stars observed with {\em K2} \citep{cody2018}, NGC~2264 had 176 stars observed with {\em CoRoT} \citep{cody2014}, and Upper Sco had 217 stars observed with {\em K2} \citep{cody2018}.

We have divided each of these datasets into disk classes based on spectral energy distribution (SED) slope, $\alpha$. Following Wilking et al.\ (2001), $\alpha = d\log(\lambda F_\lambda)/d\log(\lambda)$ for flux $F_\lambda$ as a function of wavelength $\lambda$. The slope is adopted from R20, who fit it to all available photometric datapoints between 2 and 25~$\mu$m. We consider objects with $\alpha > -1.6$ to be disk-bearing sources. Those with $-1.6 < \alpha < -0.3$ are labelled as class II, whereas SEDs with $-0.3 < \alpha < 0.3$ are flat, and those with greater slopes are class I ($\alpha > 0.3$). The majority of disks for each set are class II. 

\begin{figure}
\epsscale{1.2}
\plotone{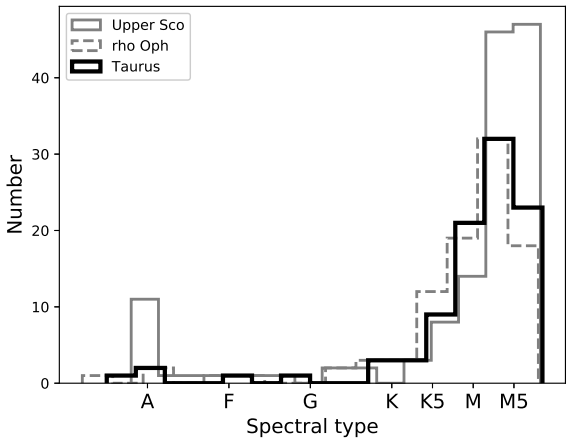}
\caption{Spectral type distributions for the young, low-mass stars in star forming regions observed with {\em K2}: $\rho$ Oph, Taurus, and Upper Sco.
Although the number of cluster members with good {\em K2} lightcurves varies among the regions, in each cluster, a similar stellar mass range is covered.}
\label{sptdist}
\end{figure}

For the Taurus sample, 70 sources are class II, 11 are flat, and four are class I. The remaining objects are class III disks candidates, with possible infrared excesses apparent on visual examination of the SED. Class I objects are the most rare, with 10 in $\rho$~Oph, four in Taurus, three in NGC~2264, and one in Upper Sco. These numbers are a reflection of the relative ages of these clusters \citep{reipurth2008}, with the $\rho$~Oph population being the youngest, on average, at $\sim$1~Myr, the Taurus and NGC 2264 regions at 1-5 Myr, and Upper Sco the oldest at $\sim$5--10~Myr.

 A histogram of the spectral type distribution for $\rho$~Oph, Taurus, and Upper Sco is shown in Figure~\ref{sptdist}.  The spectral type and, correspondingly, the mass range for each of these samples is similar; low-mass K and M stars predominate at $\sim$85--95\% of the stars. In Taurus, spectral types for all but three of our stars (EPIC~247027353, EPIC~247590222, and EPIC~248038058) have been determined by \cite{2018AJ....156..271L}; we adopt all of his values here, except for that of EPIC~247992575, which we find to be a B5 star, based on the work of \cite{2013ApJ...771..110M}.

\startlongtable
\begin{deluxetable*}{ccccc}
\tabletypesize{\scriptsize}
\tablecolumns{7}
\tablewidth{0pt}
\tablecaption{Young inner disk-bearing stars in $K2$ Campaign 2}
\tablehead{
\colhead{EPIC id} & \colhead{2MASS id} & \colhead{Other} & \colhead{Blend} & \colhead{Note} \\ 
\colhead{} & \colhead{} & \colhead{name} & \colhead{} & \colhead{} \\
}
\startdata
210683818 & J04313407+1808049 & RAFGL 5123 & - & - \\
210689083 & J04313747+1812244 & V1213 Tau & - & - \\
210690598 & J04313613+1813432 & LkHA 358 & - & - \\
210690735 & J04300399+1813493 & UX Tau ABC & Y & - \\
210690892 & J04314007+1813571 & XZ Tau AB & Y & - \\
210690913 & J04313843+1813576 & HL Tau & - & - \\
210699670 & J04315968+1821305 & - & - & - \\
210699801 & J04315779+1821350, J04315779+1821380 & V710 Tau AB & Y & - \\
210725857 & J04285053+1844361 & - & - & - \\
210777988 & J04215943+1932063 & T Tau & - & - \\
210780789 & J04221332+1934392 & - & - & M \\
210865655 & J04102834+2051507 & - & - & - \\
211104793 & J04124068+2438157 & - & - & - \\
246859790 & J04440164+1621324 & - & - & - \\
246923113 & J04470620+1658428 & DR Tau & - & - \\
246925324 & J04465305+1700001 & DQ Tau & Y & - \\
246929818 & J04465897+1702381 & Haro 6-37 & Y & - \\
246942563 & J04542368+1709534 & - & - & - \\
246963876 & J04343128+1722201 & - & - & - \\
246989752 & J04384725+1737260 & - & - & - \\
246990243 & J04384502+1737433 & - & - & - \\
247027353 & J04332789+1758436 & - & - & - \\
247031423 & J04333297+1801004 & HD 28867 AB & Y & - \\
247035365 & J04335283+1803166 & - & - & - \\
247046059 & J04324107+1809239 & RXJ0432.6+1809 & - & - \\
247047380 & J04334871+1810099 & DM Tau & - & - \\
247051861 & J04321606+1812464 & MHO 5 & - & - \\
247078342 & J04322210+1827426 & MHO 6 & - & - \\
247103541 & J04363081+1842153 & HD 285893 & - & - \\
247520207 & J04391779+2221034 & LkCa 15 & - & - \\
247534022 & J04333905+2227207 & - & - & - \\
247573157 & J04332621+2245293 & XEST 17-036 & - & - \\
247575425 & J04331907+2246342 & IRAS 04303+2240 & - & - \\
247575958 & J04330945+2246487 & CFHT-BD-Tau 12 & - & - \\
247583818 & J04354733+2250216 & HQ Tau & Y & - \\
247584113 & J04335200+2250301 & CI Tau & Y & - \\
247585465 & J04322415+2251083 & - & ? & - \\
247589612 & J04324911+2253027 & JH 112 & Y & - \\
247590222 & J04331650+2253204 & IRAS 04302+2247 & - & - \\
247591534 & J04355760+2253574 & - & - & - \\
247592103 & J04355415+2254134, J04355349+2254089 & CoKu HP Tau G2, G3 & Y & M \\
247592463 & J04355277+2254231 & HP Tau & Y & - \\
247592919 & J04355684+2254360 & Haro 6-28 & Y & - \\
247593952 & J04355209+2255039 & HP Tau/G1 & - & - \\
247596872 & J04335252+2256269, J04335243+2256323 & XEST 17-059 & Y & M \\
247601658 & J04345693+2258358 & XEST 08-003 & - & M \\
247604953 & J04305028+2300088 & Bad 16 & - & - \\
247763883 & J04330622+2409339 & GH Tau & - & - \\
247764745 & J04330664+2409549 & V807 Tau & Y & - \\
247788960 & J04323058+2419572 & FY Tau & - & - \\
247789209 & J04323176+2420029 & FZ Tau & - & - \\
247791801 & J04333456+2421058 & GK Tau & - & - \\
247792225 & J04333405+2421170 & GI Tau & - & - \\
247794636 & J04321786+2422149 & CFHT-Tau-7 & - & M \\
247799571 & J04315056+2424180 & HK Tau & Y & - \\
247805410 & J04302961+2426450 & FX Tau & - & - \\
247810494 & J04345542+2428531 & AA Tau & - & - \\
247810751 & J04321540+2428597 & Haro 6-13 & - & - \\
247820507 & J04292373+2433002 & GV Tau & - & - \\
247820821 & J04295950+2433078 & - & - & - \\
247827638 & J04293606+2435556 & XEST 13-010 & - & - \\
247837468 & J04293008+2439550 & IRAS 04264+2433 & - & - \\
247842020 & J04305171+2441475 & ZZ Tau IRS & - & - \\
247843485 & J04305137+2442222 & ZZ Tau & Y & - \\
247885481 & J05023985+2459337 & - & - & - \\
247890584 & J05010116+2501413 & - & - & - \\
247915927 & J04442713+2512164 & IRAS S04414+2506 & - & - \\
247923794 & J04423769+2515374 & DP Tau & Y & - \\
247932205 & J04403979+2519061 & - & - & - \\
247935061 & J04430309+2520187 & GO Tau & - & - \\
247935696 & J04422101+2520343 & CIDA 7 & - & - \\
247941378 & J04420548+2522562 & V999 Tau & Y & - \\
247941930 & J04420777+2523118 & LkHa 332 & Y & - \\
247968420 & J04414825+2534304 & - & - & - \\
247986526 & J04270280+2542223 & DF Tau & Y & - \\
247991214 & J04390396+2544264 & - & - & - \\
247992574 & J04392090+2545021 & GN Tau & - & - \\
247992575 & J04395574+2545020 & IRAS 0437+257P08 & - & - \\
248006676 & J04404950+2551191 & JH 223 & Y & - \\
248009353 & J04324303+2552311, J04324282+2552314 & UZ Tau AB & Y & - \\
248015397 & J04411078+2555116 & ITG 34 & - & - \\
248017479 & J04410826+2556074 & ITG 33A & - & - \\
248018164 & J04413882+2556267 & Haro 6-33 & - & - \\
248018652 & J04305718+2556394 & KPNO-Tau 7 & - & - \\
248023915 & J04380083+2558572 & ITG 2 & - & M \\
248029373 & J04304425+2601244 & DK Tau AB & Y & - \\
248029954 & J04394748+2601407 & ITG 17 & - & - \\
248030407 & J04394488+2601527 & IRAS F04366+2555 & - & - \\
248038058 & J04400800+2605253 & IRAS 04370+2559 & - & - \\
248040905 & J04295156+2606448 & IQ Tau & - & - \\
248044306 & J04300724+2608207 & KPNO-Tau 6 & - & - \\
248046139 & J04382134+2609137 & GM Tau & - & - \\
248047443 & J04333678+2609492 & IS Tau & Y & - \\
248049115 & J04383528+2610386 & HV Tau ABC & Y & - \\
248049475 & J04382858+2610494 & DO Tau & - & - \\
248051303 & J04381486+2611399 & ITG 3 & - & - \\
248055184 & J04335470+2613275 & IT Tau AB & Y & - \\
248057096 & J04331435+2614235 & IRAS S04301+2608 & - & - \\
248058354 & J04334465+2615005 & - & - & - \\
\enddata
\tablecomments{\label{tab:obstable} Stars with inner disks observed in by {\em K2} in Taurus, in order of EPIC id. In column 4, a 'Y' appears for blends-- cases in which ground-based photometry or spectroscopy indicates at least one other star contaminating the {\em K2} aperture. Objects with uncertain disk status, as determined by the spectral energy distribution analysis in \citet{rebull2020} have an 'M' in the final column.}
\end{deluxetable*}

\section{{\em K2} Photometry and Light Curves}
\label{photometry}

\begin{figure}
\epsscale{1.3}
\plotone{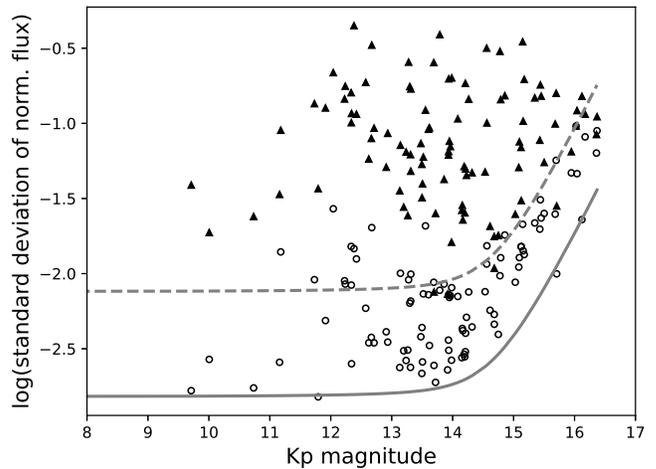}
\caption{Logarithm of the standard deviation of normalized flux for each of our targets. Open circles show the standard deviation of the fully detrended light curves from \texttt{k2sc} (i.e., both pointing systematics and variability have been largely removed), while triangles show light curves detrended only for pointing jitter (i.e., variability is preserved). The solid grey curve illustrates the expected photometric uncertainty based on Poisson photon statistics, a low-level sky background, and a systematic at the 0.15\% level. We have shifted this up by a factor of 5 to produce the dashed grey curve, which we adopt as our threshold for variability detection.}
\label{magstd}
\end{figure}

Starting with the sample described in Section~2, we obtained {\em K2} data for each disk-bearing Taurus member observed during the mission. These data were in the form of target pixel files (``TPFs''), which consist of stacks of small ($\sim$8$\times 8$ pixels or $\sim$32\arcsec$\times 32$\arcsec ) images, each taken 30 minutes apart. We performed aperture photometry on each image within a TPF, with circular aperture radii ranging from one to four pixels. Aperture placement was determined by computing a flux-weighted centroid. We then output the background subtracted summed flux. This is subject to intra-pixel positional jitter due to spacecraft drift, with corrective thruster firings approximately every six hours. The resulting photometry displays a sawtooth pattern with amplitude up to 10\%, depending on source brightness. Fortunately this systematic effect can be largely removed with detrending algorithms. To clean our light curves, we used the \texttt{k2sc} package \citep{aigrain2016} to remove position-correlated flux variations. This software simultaneously models stellar variability and pointing systematics. It also identifies flux outliers as part of this process. The outputs of \texttt{k2sc} are a ``position-dependent" trend (i.e., systematics) and a ``time-dependent" trend (i.e., real variability). We remove the former to produce our final, variability-preserved light curves, while we also subtract the latter in efforts to assess the intrinsic noise levels of the photometry, as described below and illustrated in Figure~\ref{magstd}.

We were then left with four detrended light curves (one for each aperture), along with a set of outlier flags. Each light curve was visually inspected, and an optimal aperture was chosen based on noise level and stellar crowding. For isolated stars, we picked the light curve with the lowest standard deviation after subtraction of intrinsic variability patterns (as provided by \texttt{k2sc}'s time-dependent trend output). We then checked whether the particular aperture enclosed flux from any neighboring stars; if so, we opted for a smaller aperture, typically one pixel in radius. Finally, we again used visual inspection to identify cases for which stellar flares were flagged as outliers, and we manually removed the flags on those datapoints. After outlier removal, we normalized all light curves by their median flux level. These final cleaned light curves will be available as a high level science product (HLSP) within the Mikulski Archive for Space Telescopes (MAST).

The underlying photometric noise level of our light curves is challenging to measure because of the high levels of astrophysical variability in the majority of targets. We have obtained a rough estimate of precision versus magnitude by analyzing the noise levels of the {\em variability-detrended} light curves output by \texttt{k2sc}. We estimated the {\em Kepler} magnitudes ($K_p$) by adopting the relation between raw instrumental magnitude (i.e., -2.5$\log (\rm flux)$) and $K_p$ derived for our Campaign 2 work on Upper Sco and $\rho$~Oph \citep{cody2018} based on the photometry of quiet field stars. The resulting magnitude versus noise trend is shown in Figure~\ref{magstd}. We have overplotted the standard deviations of each light curve to illustrate the significant brightness fluctuations in the majority of sources. 

Based on the variability detrended light curves, the brightest sources ($K_p\sim$10) in our sample have photometric precisions of $\sim$0.15\% in normalized flux. In contrast, fainter stars ($K_p\sim$15-16) have lower precisions of $\sim$0.65\%. In Figure~\ref{magstd}, we have shifted the noise curve up by a factor of five to generate a threshold for variability detection. All stars with standard deviations lying above this shifted curve are considered to be variable. There are then only 11 stars (black triangles in Figure~\ref{magstd}) that do not immediately qualify as variable, of which four are of uncertain disk status (`M' in Table~1). We have run these stars through periodicity detection algorithms as part of the statistics computations described in Section~4, and we find each to have one or more significant periodicities. Therefore, we are left with {\em no} non-variable objects in the entire dataset of 99 young stars. 

Because of the large {\em Kepler} pixel size ($\sim$4\arcsec ), a number of our targets are blended with companion stars. We cannot separate the photometry for most of these, and therefore have noted their status in Table~1. Blends include both visual companions and spectroscopic or photometric binary systems drawn from the literature. The light curves for blended sources contain contributions from both stars, and should therefore be considered with caution. 

We display the full set of disk bearing light curves in Appendix A.  
In each panel we show the available lightcurve, from either Campaign 4 or Campaign 13. Spatial overlap between these campaigns was minimal (two of 84 CCD channels), and so no disk bearing stars appeared in both sets. Labelling indicates the lightcurve variability class, and the quantitative $Q$ and $M$ metrics developed below.   

\begin{figure*}
\epsscale{1.15}
\plotone{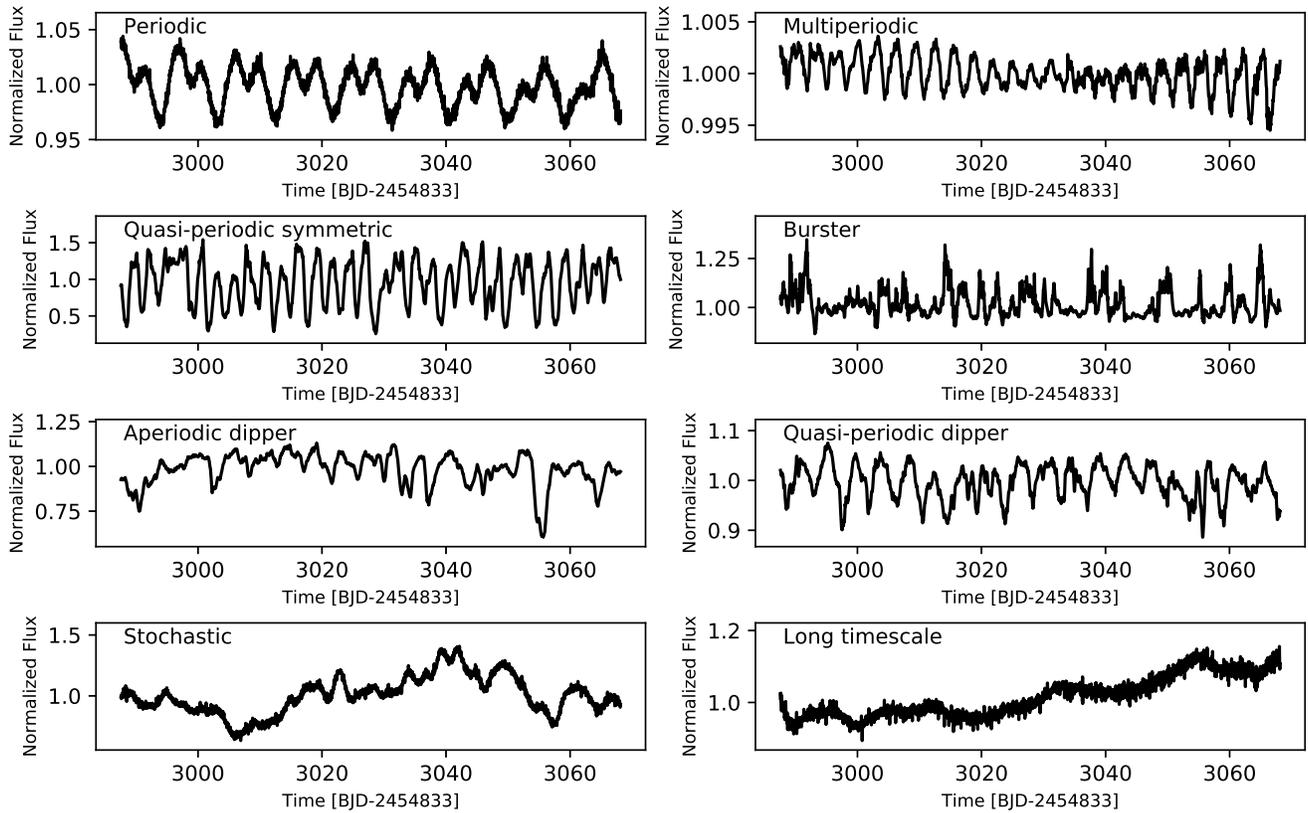}
\caption{Examples of different light curve morphologies seen among disk-bearing stars in Taurus. As with previous clusters, we have divided the sample into eight different types of behavior (excluding eclipsing binary systems, which do not appear here). All light curves are normalized by their median flux level.}
\label{examples}
\end{figure*}

\section{Variability classes} 

\subsection{$Q-M$ Classification}

Visual inspection and statistical analysis of the {\em K2} Taurus light curve sample revealed that 100\% of our inner disk-bearing stars are variable. To further probe the nature of the variability, we applied several statistical metrics introduced in our previous work on young stars observed with {\em CoRoT} and {\em Kepler}. In particular, we defined two metrics, ``$Q$'' and ``$M$'' which reflect the degrees of periodicity and flux symmetry, respectively \citep{cody2014,cody2018}. Although alternative statistics exist in the literature \citep[e.g.,][]{antille1982}, we have found that these metrics work well for capturing the range of young star variability behavior. 
Recently, they have also been applied to a ground-based data set \citep{hillenbrand2022} where extra care is needed given the larger photometric errors and lower sampling relative to space-based photometry.

$Q$ values range from zero to one, where $Q=0$ indicates a perfectly periodic light curve, and $Q=1$ is obtained for a light curve with no repeating patterns. Intermediate values of $Q$ apply to flux behavior that is somewhat repetitive but does not display the same shape or amplitude from one cycle to the next. The $M$ statistic is a measure of light curve symmetry, where $M=0$ is retrieved for a light curve with flux equally distributed above and below the median. Positive $M$ values indicate a tendency toward fading events, whereas negative values indicate predominantly brightening behavior. For more discussion and the exact definition of $Q$ and $M$, we refer the reader to \citet{cody2014} and \citet{cody2018}.

\begin{figure}
\epsscale{1.3}
\plotone{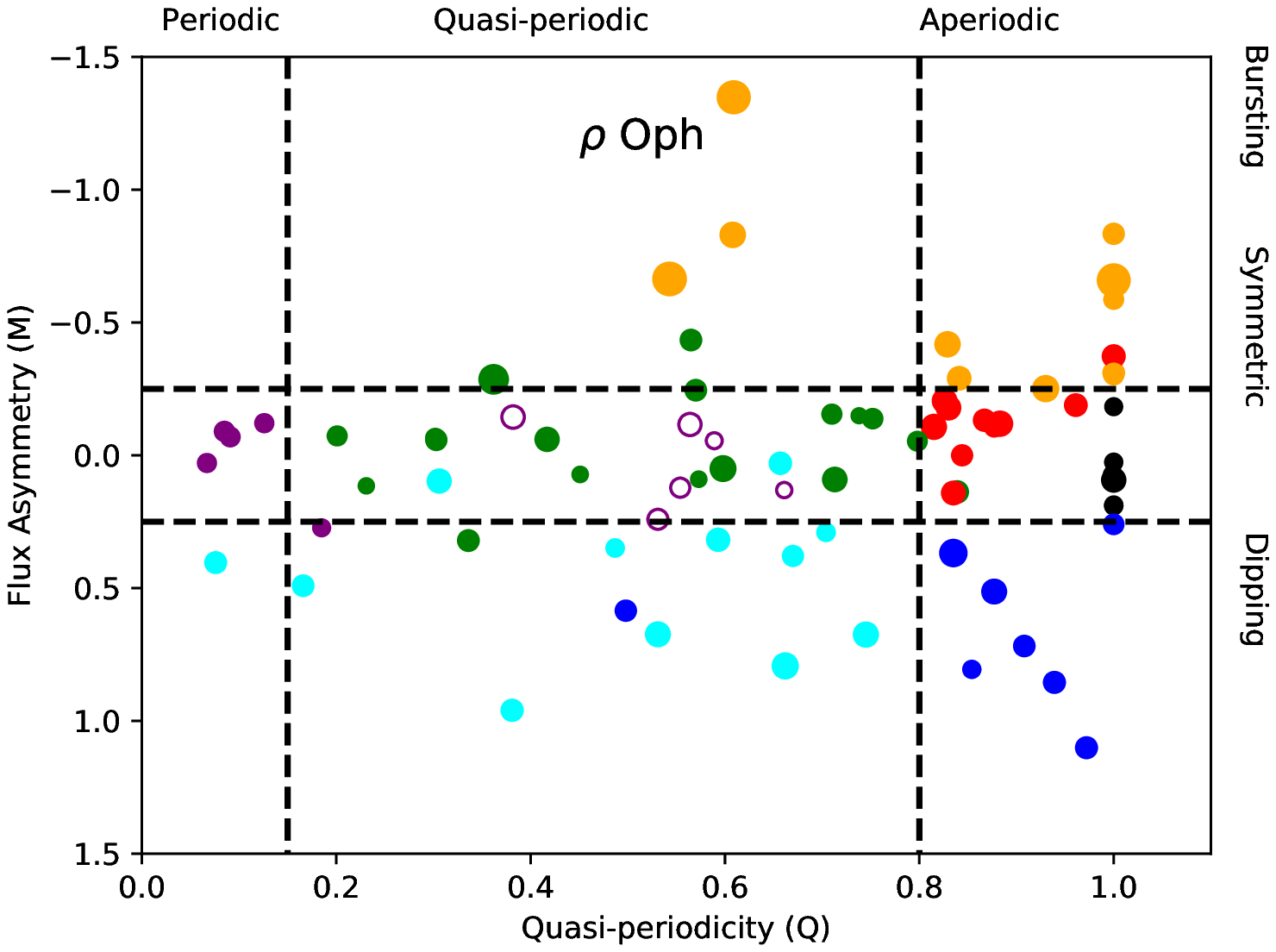}
\plotone{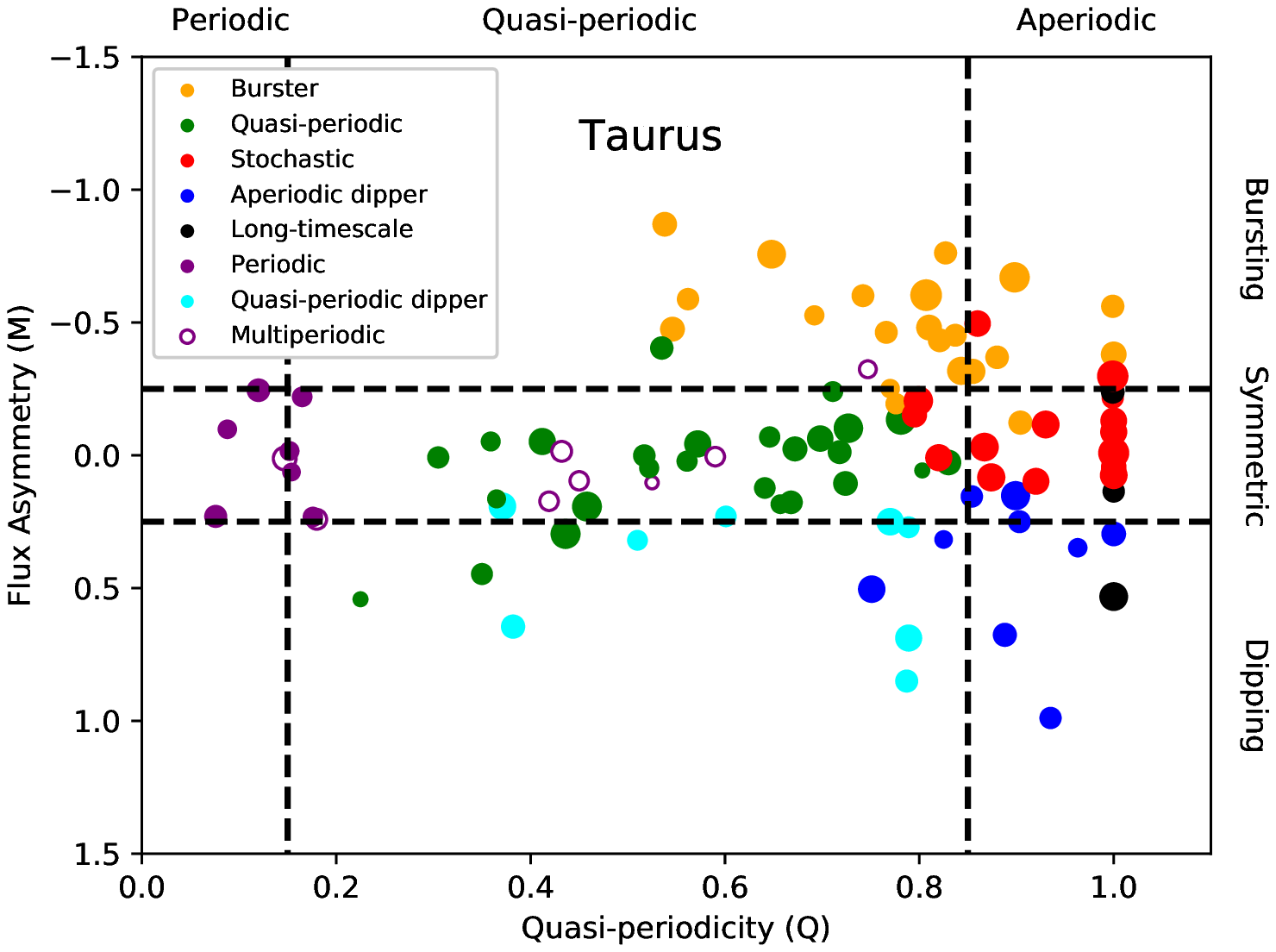}
\plotone{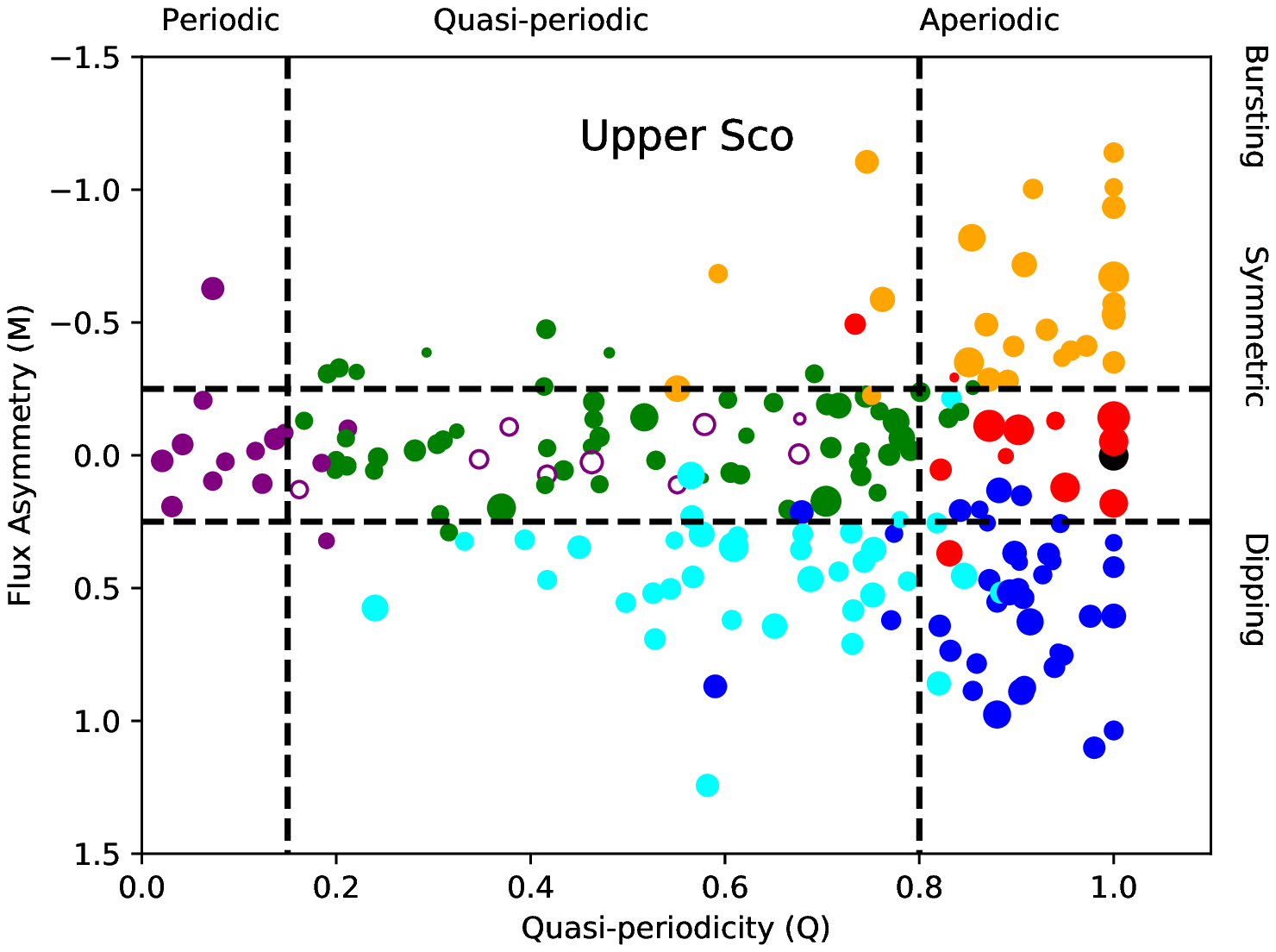}
\caption{$Q$ and $M$ statistics for our sample of disk-bearing stars in $\rho$~Ophiuchus (top), Taurus (middle), and Upper Scorpius (bottom). Non-variable objects are excluded.  Point areas in this and subsequent plots are scaled according to variability amplitude to the one-third power (see \S5.1).  Colors in this and subsequent plots denote different types of variables, as identified by eye; see text. 
}
\label{qmplot}
\end{figure}

Using the derived statistics, we have classified the disk-bearing variables of Taurus into eight different categories, as defined in \cite{cody2014} and shown in Figure~\ref{examples}. In accordance with our analysis of other {\em K2} young star light curves \citep{cody2018}, the categories, their shorthand denotation, and their $Q$ and $M$ ranges are: 

\begin{itemize}
\item burster (B) \\ $M<-0.25$
\item purely periodic symmetric (P) \\ $Q<0.15$ and $-0.25<M<0.25$
\item quasi-periodic symmetric (QPS) \\ $0.15<Q<0.85$ and $-0.25<M<0.25$ 
\item purely stochastic (S) \\ $Q>0.85$ and $-0.25<M<0.25$
\item quasi-periodic dipper (QPD) \\$0.15<Q<0.85$ and $M>0.25$
\item aperiodic dipper (APD) \\$Q>0.85$ and $M>0.25$
\item long timescale (L)
\item variable but unclassifiable (U).  
\end{itemize}

The first six categories are the same as those used in our other {\em K2} young star analyses in the $\rho$~Oph/Upper Sco region \citep{cody2018} and in NGC~6530, the Lagoon Nebula cluster \citep{venuti2021}. 
We note that \cite{venuti2021} additionally included categories for multi-periodic and eclipsing binary light curves, which we do not find here.
Our final two categories are reserved for ``long-term" variables that showed a trend on $>30$ day timescales, as well as ``unclassifiable" sources with unknown light curve form but high enough standard deviations to be selected as variable. As noted in Section~\ref{photometry}, we do not have any non-variable sources in this sample. 

The distribution of $Q$ and $M$ values for all disk-bearing Taurus stars observed by {\em K2}, as well as those in $\rho$~Oph and Upper Scorpius, is displayed in Figure~\ref{qmplot}. In those plots, we illustrate the established divisions in $Q$ and $M$, but emphasize that these should be viewed moreso as transition regions from one variability type to another, rather than as sharp boundaries.

In Taurus, the $Q$ distribution spans the whole range from zero to one. The results for the $M$ statistic indicate that 40\% of Taurus stars exhibit asymmetric flux behavior. tending to either preferentially fade or to brighten. There are no eclipsing binary systems in our sample; instead, the fading events are associated with dippers (17\% prevalence). The remaining 23\% of asymmetric flux sources are bursters. 

We list the variability status of all disk-bearing stars (illustrated in Appendix A) in Table~2, including the metrics regarding the degree of periodicity and flux asymmetry about the mean value,  the inferred variability type, and the lightcurve amplitude and its timescale. 
The derivation of the last two quantities is discussed in \S5 below. 

\subsection{Hybrid Classes}

Table 2 indicates 10 sources with multiple variability categories. Notably, within the wide phase-space from periodic, to quasi-periodic in $Q$, to stochastic, and that from dipper, to symmetric in $M$, to burster, only a small part of the $Q-M$ classification continuum is populated by the group of stars having hybrid light curve morphologies. Of these 10 sources, 8 are a combination of the quasi-periodic-symmetric (QPS) class with the aperiodic dipper (APD) class, one is a combined QPS and burster (B), and one is a combined periodic (P) and APD.

We have cross-examined the Table 2 hybrids with the Table 1 blends, so as to determine if the multi-faceted light curve behavior is the result of stellar multiplicity.  For most sources there is no indication of binarity in previous multiplicity studies of the Taurus population.   The two exceptions are  EPIC 246925324 (DQ Tau) which is a known spectroscopic binary and JH 112 which are both discussed in Appendix B.

Another set of objects to note is those having $Q$-$M$ behavior that changes over the nearly 3-month duration of the $K2$ observations. An example is EPIC~204538777 with stochastic or otherwise noisy behavior over a period of several weeks, after which dipping suddenly appears to turn on.  These types of schizophrenic young stars have been discussed in more detail by e.g. \cite{mcginnis2015} who highlight changes between (quasi-)periodic and aperiodic behavior on timescales of years.

\subsection{Summary of Source Classification}

Table 3 shows the breakdown of the Taurus classification.  We find 21 bursters, 16 stochastic objects, nine aperiodic dippers, eight quasi-periodic dippers, 26 quasi-periodic symmetric objects, seven periodic stars, eight multiperiodic sources, and three long-timescale variables. Uncertainties on the fractional make-up for each group were determined based on binomial distributions \citep[see, e.g., the appendix of][]{2003ApJ...586..512B}. We also compare in Table 3 the Taurus fractions to those for other clusters we have studied, and discuss these results in \S7.

\section{Variability Properties}

In addition to computing the $Q$ and $M$ statistics for each light curve, we have also assessed other features such as the variability amplitude and various timescale metrics. These measurements provide a broader picture of the flux variations, offering hints into their origins. 

\begin{figure}
\epsscale{1.3}
\plotone{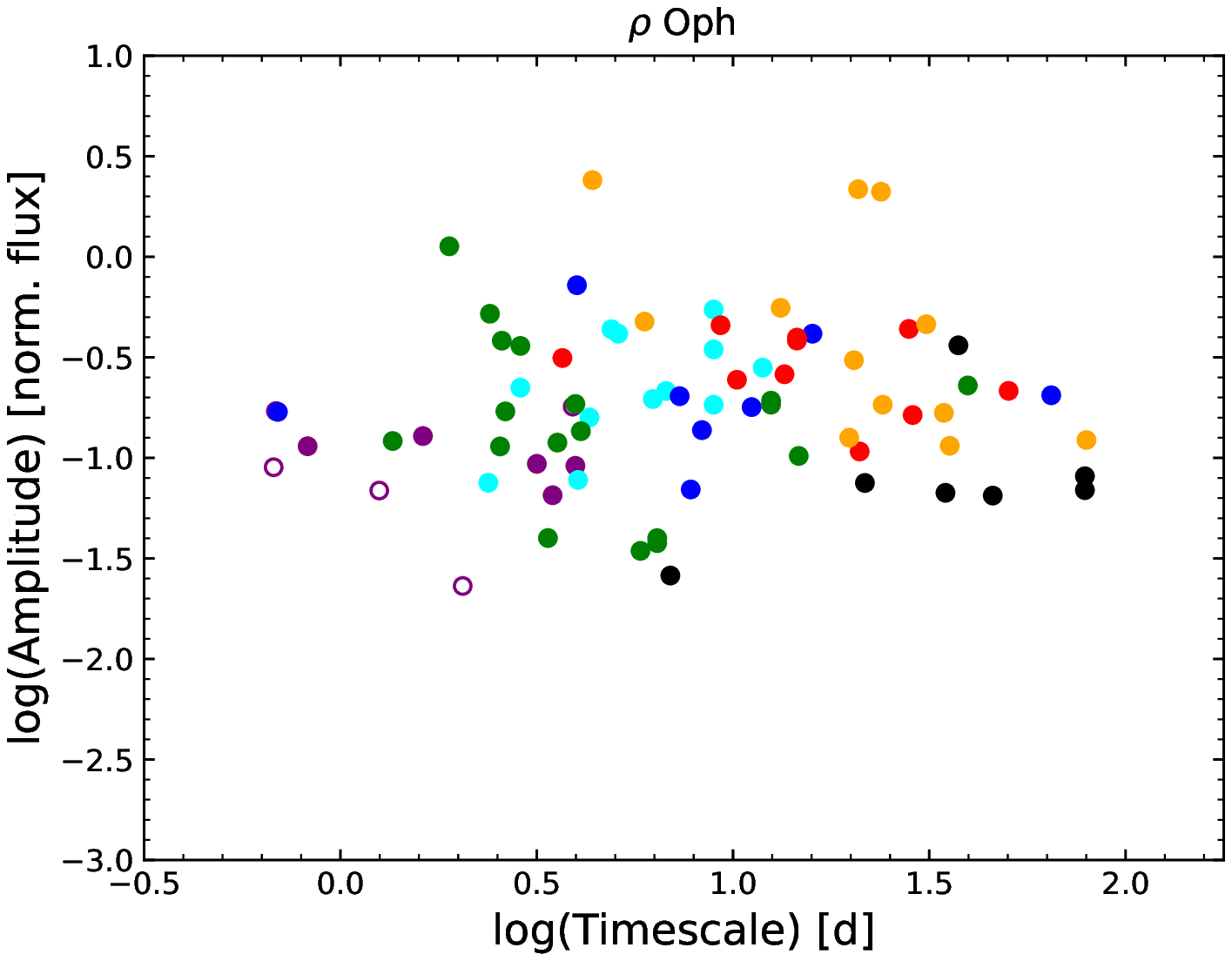}
\plotone{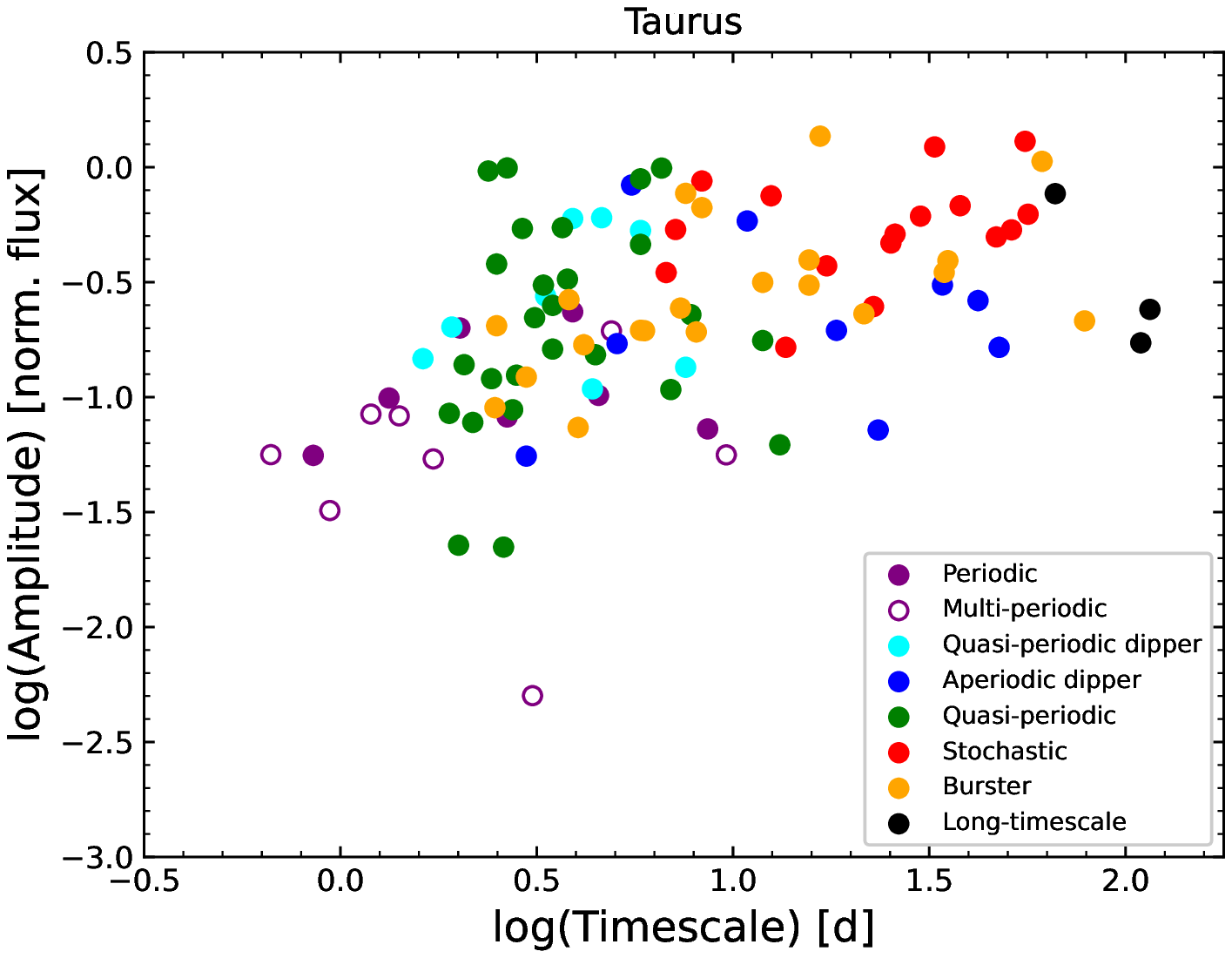}
\plotone{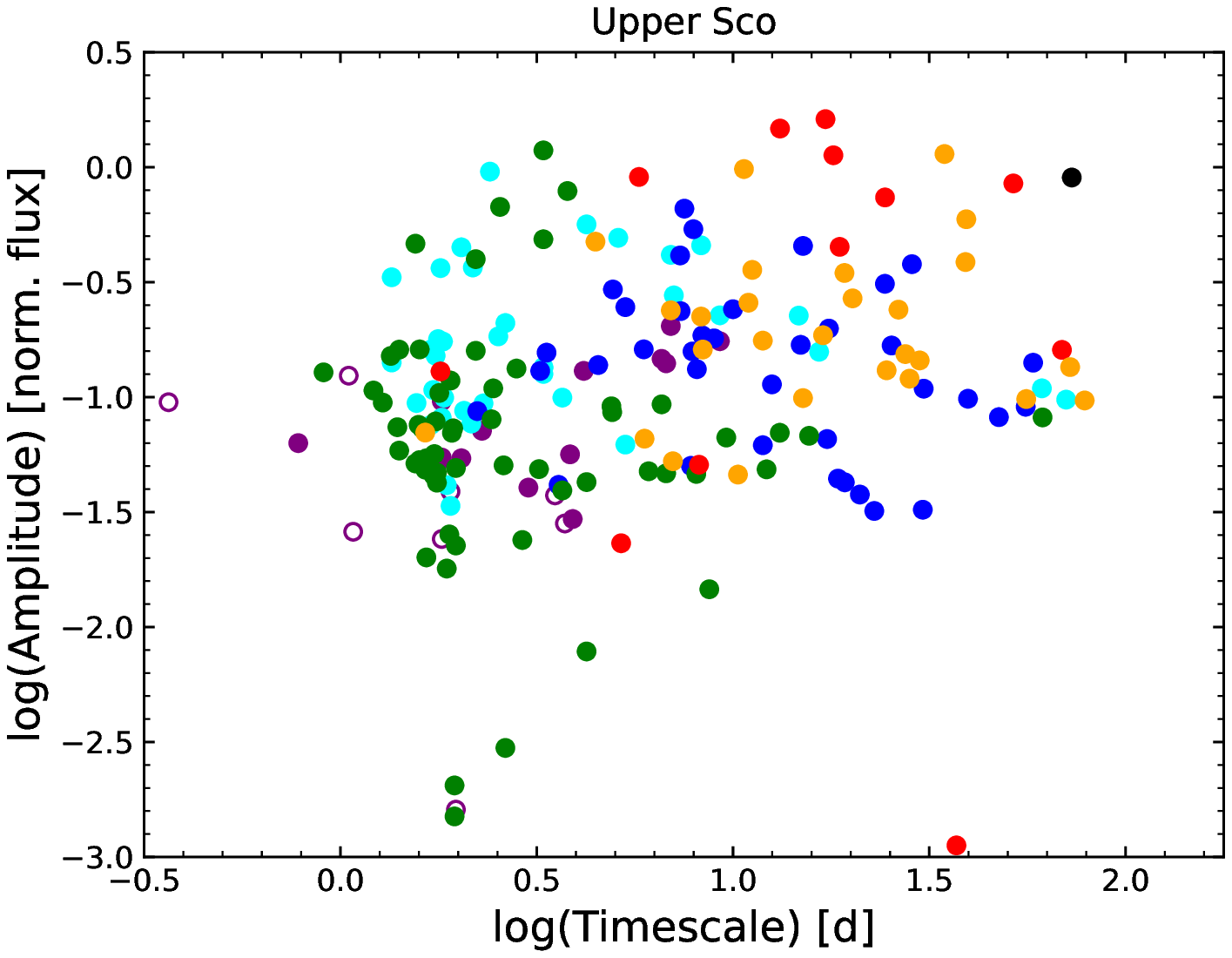}
\caption{Amplitudes and timescales measured for the samples of disk-bearing YSOs in $\rho$~Oph (top), Taurus (middle) and Upper Sco (bottom). Each point is colored by its variability type; no size scaling is applied here, since amplitude is one of the axes. See text for discussion.}
\label{amptimescale}
\end{figure}

\subsection{Amplitudes}

While the photometric precision of most of our light curves is less than 1\%, the observed flux excursions are typically much larger than that. To quantify the variability level and search for correlations with light curve morphology type, we measured a ``peak-to-peak" amplitude. This metric was defined by first identifying the bottom and top fifth percentile of flux points, and then determining the difference between them. All amplitudes were measured on the normalized light curves. The resulting values are reported in Table~2 and plotted in Figure~\ref{amptimescale}.

The variability amplitudes range from 0.005 to 1.36 in normalized flux units (roughly 0.5\% to 140\%). 
As seen in Figure~\ref{amptimescale}, the lowest amplitude objects are periodic, multi-periodic, or quasi-periodic, 
while those with the highest amplitudes tend to be aperiodic, though not exclusively so.

We note that the point sizes in Figures~\ref{qmplot}, \ref{Halpha_alpha}, \ref{taurusinclinations}, and \ref{alphaspt} are scaled by these lightcurve amplitudes. 

\subsection{Timescales}

The timescale of light curve variations is challenging to quantify, but is of interest for the assessment of variability origins and correlations with [circum]stellar properties. We start with the relatively easy case of repeating flux patterns and progress to more complex light curves with predominantly stochastic behavior.

\subsubsection{Periodicities and Quasiperiodicities}

In the few cases for which a light curve displays perfectly periodic behavior, the timescale is straightforward to identify; it appears as a prominent peak in the autocorrelation function and periodogram. We initially calculate the former and identify the tallest peak. The associated autocorrelation timescale may not be very precise, so we refine it by identifying the corresponding peak in a Fourier transform periodogram and determining the frequency. We then assess whether this periodogram peak is significant. To do so, we evaluate its height with respect to the underlying noise trend. If the amplitude of the peak is at least four times the periodogram noise level, then we consider it significant \citep[e.g.,][]{1993A&A...271..482B}. We also use the $Q$ statistic itself as a measure of periodicity; based on our previous work, $Q$ values less than 0.15 are indicative of highly periodic sources. 

Many stars in our sample exhibit repeating flux patterns that change in amplitude or shape from one cycle to the next. For these cases, we can still attach a timescale to the light curve through examination of the periodogram and autocorrelation function. Both of these were calculated in the determination of $Q$. We again take any significant periodogram peak to indicate an intrinsic variability timescale, but we consider the star quasi-periodic if $Q$ lies between 0.15 and 0.85. In a few cases, this label can be overridden, e.g., if the source is clearly multiperiodic and imperfect removal of the first period drives the $Q$ value upward.  Most of the light curves with clear timescales are relatively symmetric in terms of their flux excursions ($-0.25 < M < 0.25$), but some bursters and dippers have quasiperiodic behavior; none have strictly periodic behavior.

\subsubsection{Charaterizing Aperiodic Behavior}
For targets with no periodic or quasi-periodic signals identified in the light curve, we can still attach a rough timescale to the variations. We employ the PeakFind algorithm outlined in \citet{cody2014} and \cite{Findeisen2015}. For a given amplitude (ranging from the light curve noise floor to its peak-to-peak maximum described in Section~5.1.1), PeakFind locates the local maxima and minima that differ by at least that value. The median spacing between these maxima and minima then equate to a timescale. We double this number so that it equals the period when the input light curve is a sinusoid. 

PeakFind results in a series of timescales that are a function of amplitude. While most variability is not characterized by a single timescale, we can associate a value to the ``typical'' variations. Past work \citep[e.g.,][]{cody2014} has suggested that the right ``typical" timescale to use is 70\% of the peak-to-peak amplitude, as representative of the behavior. We therefore use this value to attach an aperiodic timescale to each light curve, as specified by PeakFind. 

We note that this method requires sampling of the detailed inflections in lightcurves and would not work on more sparsely sampled ground-based data sets for YSOs. The timescales computed for all Taurus objects, whether periodic or aperiodic, are listed in Table~2 and plotted in Figure~\ref{amptimescale}. If a light curve has more than one period, we list the one with the highest autocorrelation peak first; this is also the timescale adopted for Figure~\ref{amptimescale}. As with previously studied clusters, we find that the [multi-]periodic objects display the shortest timescales (approximately one to a few days), while the stochastic and long-timescale stars take longer to vary (typically at least 10 days) at the most dominant amplitudes. Quasi-periodic stars tend to display patterns repeating on timescales of a few days to a week, while the bursters brighten slightly less frequently and over a larger range of timescales (three days to three weeks).


\startlongtable
\begin{deluxetable*}{ccccccccc}
\tabletypesize{\scriptsize}
\tablecolumns{9}
\tablewidth{0pt}
\tablecaption{Variability properties of young disk
bearing stars in $K2$ Campaign 2}
\tablehead{
\colhead{EPIC} & \colhead{2MASS} & \colhead{Variability} & \colhead{Variability} & \colhead{Amplitude} & \colhead{Timescale} & \colhead{Additional} & \colhead{$Q$} & \colhead{$M$}\\ 
\colhead{id} & \colhead{id} & \colhead{(primary)} & \colhead{(secondary)} &  \colhead{(Norm. Flux)} & \colhead{(d)} & \colhead{period (d)} & \colhead{} & \colhead{}\\
}
\startdata
210683818 & J04313407+1808049 & L &  & 0.17 & 109.31 &  &  1.0 & 0.14 \\
210689083 & J04313747+1812244 & B &  &  0.77 & 7.58 &  &  0.65 & -0.76 \\
210690598 & J04313613+1813432 & S &  &  0.47 & 25.25 &  &  0.86 & -0.5 \\
210690735 & J04300399+1813493 & APD &  &  0.84 & 5.52 &  &  0.9 & 0.15 \\
210690892 & J04314007+1813571 & S &  &  0.5 & 46.85 &  &  1.0 & -0.13 \\
210690913 & J04313843+1813576 & S &  &  0.25 & 22.80 &  &  1.0 & -0.3 \\
210699670 & J04315968+1821305 & B &  &  1.36 & 16.67 &  &  0.81 & -0.6 \\
210699801 & J04315779+1821380 & B &  &  0.07 & 4.03 &  &  0.77 & -0.25 \\
210725857 & J04285053+1844361 & QPS & APD &  0.14 & 2.07 &  &  0.64 & 0.12 \\
210777988 & J04215943+1932063 & QPS &  &  0.12 & 2.81 &  &  0.65 & -0.07 \\
210780789 & J04221332+1934392 & P &  &  0.1 & 1.33 &  &  0.18 & 0.23 \\
210865655 & J04102834+2051507 & B &  &  0.19 & 8.06 &  &  0.74 & -0.6 \\
211104793 & J04124068+2438157 & B &  &  0.2 & 5.81 &  &  0.77 & -0.46 \\
246859790 & J04440164+1621324 & QPS & APD &  0.08 & 2.17 &  &  0.66 & 0.18 \\
246923113 & J04470620+1658428 & S &  &  0.75 & 12.50 &  &  0.87 & -0.03 \\
246925324 & J04465305+1700001 & B & QPS &  0.31 & 15.62 & 3.01 &  0.54 & -0.87 \\
246929818 & J04465897+1702381 & APD &  &  0.58 & 10.87 &  &  0.75 & 0.5 \\
246942563 & J04542368+1709534 & APD &  &  0.17 & 5.07 &  &  0.85 & 0.16 \\
246963876 & J04343128+1722201 & B &  &  0.21 & 78.62 &  &  1.0 & -0.56 \\
246989752 & J04384725+1737260 & QPD &  &  0.15 & 1.62 &  &  0.79 & 0.27 \\
246990243 & J04384502+1737433 & B &  &  0.2 & 2.50 &  &  0.83 & -0.76 \\
247027353 & J04332789+1758436 & B &  &  0.67 & 8.33 &  &  0.84 & -0.32 \\
247031423 & J04333297+1801004 & MP &  &  0.01 & 3.09 & 3.29 &  0.52 & 0.1 \\
247035365 & J04335283+1803166 & QPS &  &  0.02 & 2.60 &  &  0.22 & 0.54 \\
247046059 & J04324107+1809239 & QPS &  &  0.02 & 2.00 &  &  0.8 & 0.06 \\
247047380 & J04334871+1810099 & B &  &  0.24 & 7.35 &  &  0.82 & -0.43 \\
247051861 & J04321606+1812464 & P &  &  0.08 & 2.66 &  &  0.15 & -0.02 \\
247078342 & J04322210+1827426 & QPD &  &  0.2 & 1.92 &  &  0.79 & 0.85 \\
247103541 & J04363081+1842153 & APD &  &  0.07 & 23.44 &  &  0.96 & 0.35 \\
247520207 & J04391779+2221034 & QPD &  &  0.53 & 5.81 &  &  0.79 & 0.69 \\
247534022 & J04333905+2227207 & QPS &  &  0.06 & 13.16 &  &  0.36 & 0.16 \\
247573157 & J04332621+2245293 & MP &  &  0.06 & 9.62 & 4.55 &  0.42 & 0.17 \\
247575425 & J04331907+2246342 & S &  &  0.68 & 37.90 &  &  0.87 & 0.08 \\
247575958 & J04330945+2246487 & QPS &  &  0.25 & 3.47 &  &  0.72 & -0.01 \\
247583818 & J04354733+2250216 & QPS & APD &  0.12 & 2.43 &  &  0.56 & 0.02 \\
247584113 & J04335200+2250301 & S &  &  0.37 & 17.33 &  &  1.0 & 0.04 \\
247585465 & J04322415+2251083 & B &  &  0.09 & 2.48 &  &  0.69 & -0.53 \\
247589612 & J04324911+2253027 & APD & QPS &  0.16 & 47.65 & 2.21 &  0.94 & 0.99 \\
247590222 & J04331650+2253204 & S &  &  0.53 & 51.22 &  &  1.0 & -0.09 \\
247591534 & J04355760+2253574 & QPS &  &  0.23 & 7.81 &  &  0.67 & 0.18 \\
247592103 & J04355415+2254134 & MP &  &  0.08 & 1.20 & 1.22 &  0.43 & -0.02 \\
247592463 & J04355277+2254231 & APD &  &  0.26 & 42.09 &  &  0.89 & 0.68 \\
247592919 & J04355684+2254360 & B &  &  0.23 & 21.56 &  &  0.88 & -0.37 \\
247593952 & J04355209+2255039 & P &  &  0.06 & 0.85 &  &  0.15 & 0.06 \\
247596872 & J04335252+2256269 & MP &  &  0.08 & 1.41 & 4.10 &  0.18 & 0.24 \\
247601658 & J04345693+2258358 & P &  &  0.2 & 2.02 &  &  0.08 & 0.23 \\
247604953 & J04305028+2300088 & S &  &  0.62 & 56.45 &  &  0.93 & -0.12 \\
247763883 & J04330622+2409339 & QPS &  &  0.38 & 2.50 &  &  0.83 & 0.03 \\
247764745 & J04330664+2409549 & QPD &  &  0.11 & 4.39 &  &  0.51 & 0.32 \\
247788960 & J04323058+2419572 & QPS &  &  0.11 & 6.94 &  &  0.71 & -0.24 \\
247789209 & J04323176+2420029 & B &  &  0.39 & 35.27 &  &  1.0 & -0.38 \\
247791801 & J04333456+2421058 & QPD &  &  0.6 & 4.63 &  &  0.37 & 0.19 \\
247792225 & J04333405+2421170 & S &  &  0.54 & 7.14 &  &  0.82 & 0.01 \\
247794636 & J04321786+2422149 & MP &  &  0.05 & 1.72 & 2.60 &  0.45 & 0.1 \\
247799571 & J04315056+2424180 & QPS &  &  0.31 & 3.29 &  &  0.72 & 0.11 \\
247805410 & J04302961+2426450 & APD &  &  0.31 & 34.16 &  &  1.0 & 0.3 \\
247810494 & J04345542+2428531 & S &  &  1.3 & 55.47 &  &  1.0 & -0.3 \\
247810751 & J04321540+2428597 & S &  &  0.51 & 25.87 &  &  0.92 & 0.1 \\
247820507 & J04292373+2433002 & S &  &  0.16 & 13.63 &  &  1.0 & -0.22 \\
247820821 & J04295950+2433078 & QPS &  &  0.96 & 2.38 &  &  0.46 & 0.19 \\
247827638 & J04293606+2435556 & P & APD &  0.23 & 3.91 &  &  0.12 & -0.24 \\
247837468 & J04293008+2439550 & B &  &  0.32 & 11.90 &  &  0.55 & -0.48 \\
247842020 & J04305171+2441475 & L &  &  0.24 & 115.36 &  &  1.0 & -0.24 \\
247843485 & J04305137+2442222 & B &  &  0.17 & 4.17 &  &  0.56 & -0.59 \\
247885481 & J05023985+2459337 & APD & QPS & 0.06 & 2.98 &  &  0.82 & 0.32 \\
247890584 & J05010116+2501413 & P &  &  0.07 & 8.62 &  &  0.09 & -0.1 \\
247915927 & J04442713+2512164 & QPS & APD & 0.15 & 4.46 &  &  0.35 & 0.45 \\
247923794 & J04423769+2515374 & QPS &  &  0.55 & 3.68 &  &  0.57 & -0.04 \\
247932205 & J04403979+2519061 & MP &  &  0.03 & 0.94 & 0.85 &  0.74 & -0.32 \\
247935061 & J04430309+2520187 & APD &  &  0.2 & 18.35 &  &  0.9 & 0.25 \\
247935696 & J04422101+2520343 & QPS &  &  0.08 & 1.89 &  &  0.36 & -0.05 \\
247941378 & J04420548+2522562 & MP &  &  0.19 & 4.90 & 2.66 &  0.15 & 0.01 \\
247941930 & J04420777+2523118 & S &  &  0.87 & 8.33 &  &  0.8 & -0.2 \\
247968420 & J04414825+2534304 & QPS &  &  0.54 & 2.91 &  &  0.41 & -0.05 \\
247986526 & J04270280+2542223 & B &  &  0.4 & 15.62 &  &  0.81 & -0.48 \\
247991214 & J04390396+2544264 & QPS &  &  0.22 & 3.12 &  &  0.54 & -0.4 \\
247992574 & J04392090+2545021 & QPS &  &  0.89 & 5.81 &  &  0.73 & -0.1 \\
247992575 & J04395574+2545020 & L &  &  0.77 & 66.20 &  &  1.0 & 0.53 \\
248006676 & J04404950+2551191 & QPD &  &  0.27 & 3.33 &  &  0.38 & 0.64 \\
248009353 & J04324282+2552314 & QPS &  &  0.33 & 3.79 &  &  0.67 & -0.02 \\
248015397 & J04411078+2555116 & QPD &  &  0.6 & 3.91 &  &  0.77 & 0.25 \\
248017479 & J04410826+2556074 & S &  &  0.35 & 6.76 &  &  0.8 & -0.15 \\
248018164 & J04413882+2556267 & S &  &  0.61 & 30.03 &  &  1.0 & 0.08 \\
248018652 & J04305718+2556394 & U &  &  0.25 & 27.87 &  &  0.96 & -0.1 \\
248023915 & J04380083+2558572 & MP &  &  0.06 & 0.66 &  &  1.0 & 0.0 \\
248029373 & J04304425+2601244 & S &  &  1.23 & 32.63 &  &  1.0 & -0.01 \\
248029954 & J04394748+2601407 & B &  &  0.12 & 2.98 &  &  0.78 & -0.19 \\
248030407 & J04394488+2601527 & QPS & APD &  0.16 & 3.47 &  &  0.3 & 0.01 \\
248038058 & J04400800+2605253 & B &  &  0.35 & 34.53 &  &  0.86 & -0.32 \\
248040905 & J04295156+2606448 & QPS & APD &  0.99 & 6.58 &  &  0.78 & -0.13 \\
248044306 & J04300724+2608207 & B &  &  0.27 & 3.82 &  &  0.9 & -0.12 \\
248046139 & J04382134+2609137 & QPS &  &  0.99 & 2.66 &  &  0.44 & 0.3 \\
248047443 & J04333678+2609492 & QPD &  &  0.13 & 7.58 &  &  0.6 & 0.23 \\
248049115 & J04383528+2610386 & P &  &  0.1 & 4.54 &  &  0.16 & -0.22 \\
248049475 & J04382858+2610494 & B &  &  1.06 & 61.27 &  &  0.9 & -0.67 \\
248051303 & J04381486+2611399 & B &  &  0.19 & 5.95 &  &  0.84 & -0.45 \\
248055184 & J04335470+2613275 & QPS &  &  0.09 & 2.75 &  &  0.52 & 0.05 \\
248057096 & J04331435+2614235 & QPS &  &  0.18 & 11.9 &  &  0.52 & 0.0 \\
248058354 & J04334465+2615005 & QPS &  &  0.46 & 5.81 &  &  0.7 & -0.06 \\
\enddata
\tablecomments{\label{tab:vartable} Variability properties for Taurus stars with inner disks observed in {\em K2} Campaigns 4 and 13. Variability types are determined by eye and supported by statistical measures. The types consist of the following: ``P" is for strictly periodic behavior, ``MP" is reserved for stars with multiple distinct periods, ``QPD" is for quasi-periodic dippers, ``QPS" means quasi-periodic symmetric (i.e., quasi-periodic stars that neither burst nor dip), ``APD" are aperiodic dippers, ``B" is for bursters, ``S" is for stochastic stars, ``L" is the label for long-timescale behavior that doesn't fall into the other categories.}
\end{deluxetable*}

\begin{deluxetable*}{cccccc}
\vspace{0.8cm}
\tabletypesize{\scriptsize}
\tablecolumns{6}
\tablewidth{0pt}
\tablecaption{Variability types among young disk-bearing stars}
\tablehead{
\colhead{Morphology class} & \colhead{Taurus} & \colhead{Oph} & \colhead{Sco} & \colhead{Sco/Oph} & \colhead{NGC 2264} \\ [-0.4cm]
\colhead{} & \colhead{} & \colhead{} & \colhead{} & \colhead{composite} & \colhead{} \\ [-0.3cm]
\colhead{} & \colhead{\%} & \colhead{\%} & \colhead{\%} & \colhead{\%} & \colhead{\%} \\ [-0.4cm]
}
\startdata
\multicolumn{6}{c}{Categories based on periodicity and stochasticity}\\
\tableline
All Bursters & 21$^{+5}_{-3}$ & 14$^{+5}_{-3}$ & 13$^{+3}_{-2}$ & 14$^{+2}_{-2}$ & 13$^{+3}_{-2}$ \\
Aperiodic symmetric (Stochastic) & 16$^{+5}_{-3}$ & 12$^{+4}_{-3}$ & 6$^{+2}_{-1}$ & 8$^{+2}_{-2}$ & 13$^{+3}_{-2}$ \\
Quasi-periodic symmetric & 26$^{+5}_{-4}$ & 20$^{+5}_{-4}$ & 29$^{+3}_{-3}$ & 26$^{+3}_{-2}$ & 17$\pm$3 \\
Aperiodic dippers & 9$^{+4}_{-2}$ & 9$^{+5}_{-2}$ & 18$^{+3}_{-2}$ & 16$^{+2}_{-2}$ & 11$^{+3}_{-2}$ \\
Quasi-periodic dippers & 8$^{+4}_{-2}$ & 14$^{+5}_{-3}$ & 18$^{+3}_{-2}$ & 17$^{+2}_{-2}$ & 10.5$^{+3}_{-2}$\\
Periodic symmetric & 7$^{+4}_{-2}$ & 6$^{+4}_{-2}$ & 7$^{+2}_{-2}$ & 7$^{+1}_{-2}$ & 3$^{+2}_{-1}$ \\
\multicolumn{6}{c}{Other Categories}\\
\tableline
Multiperiodic & 8$^{+4}_{-2}$ & 7$^{+4}_{-2}$ & 4$^{+2}_{-1}$ & 5$^{+2}_{-1}$ & 1$^{+2}_{-1}$ \\
Long timescale & 3$^{+3}_{-1}$ & 8$^{+4}_{-2}$ & 0$^{+2}_{-0}$ & 3$^{+1}_{-1}$ & 1$^{+2}_{-1}$ \\
Unclassifiable & 0$^{+2}_{-0}$ & 2$^{+3}_{-0}$ & 0$^{+2}_{-0}$ & 1$^{+1}_{-1}$ & 11$^{+3}_{-2}$ \\
Non-variable & 0$^{+2}_{-0}$ & 6$^{+4}_{-2}$ & 3$^{+2}_{-1}$ & 4$^{+1}_{-1}$ & 19$\pm$3 \\
\enddata
\tablecomments{\label{tab:varfractions} Fraction of young stars in each light curve morphology group, as defined by eye but generally supported by the statistical measures $Q$ and $M$ (\S5). There are eight variability categories, plus one more for non-variable sources.}
\end{deluxetable*}

\section{Connection between variability and stellar/circumstellar properties}

\subsection{Variability and Circumstellar Disks}

As described in \S2, our sample consists of 99 disk-bearing sources.  Of these, 93 have evidence for strong inner disks with $\alpha > -1.6$. The remainder have weaker disks with
$\alpha < -1.6$ (where a photosphere would have $\alpha = -3$); they are formally Class III objects interpreted as having sparsely populated inner regions.

\begin{figure}
\epsscale{1.3}
\plotone{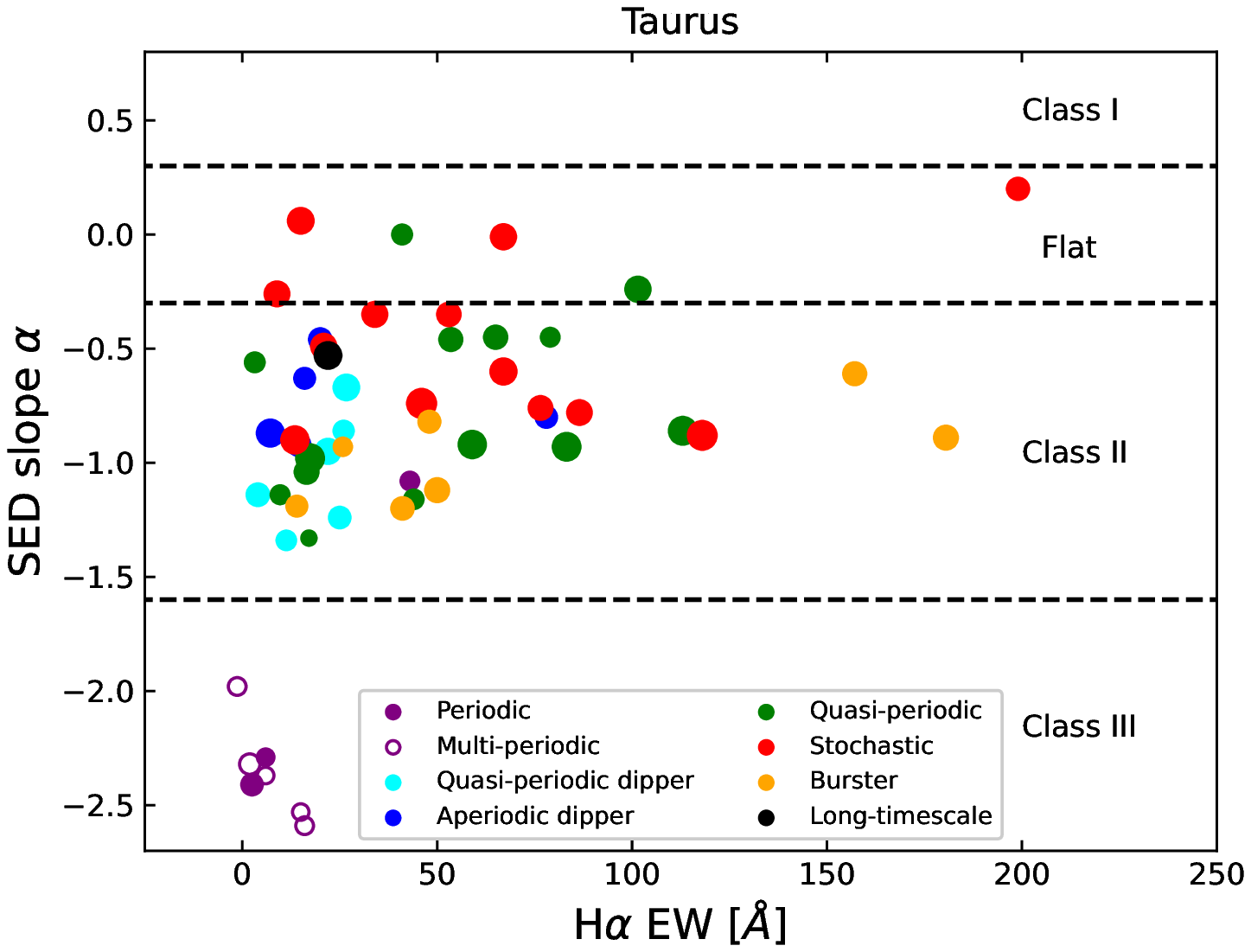}
\plotone{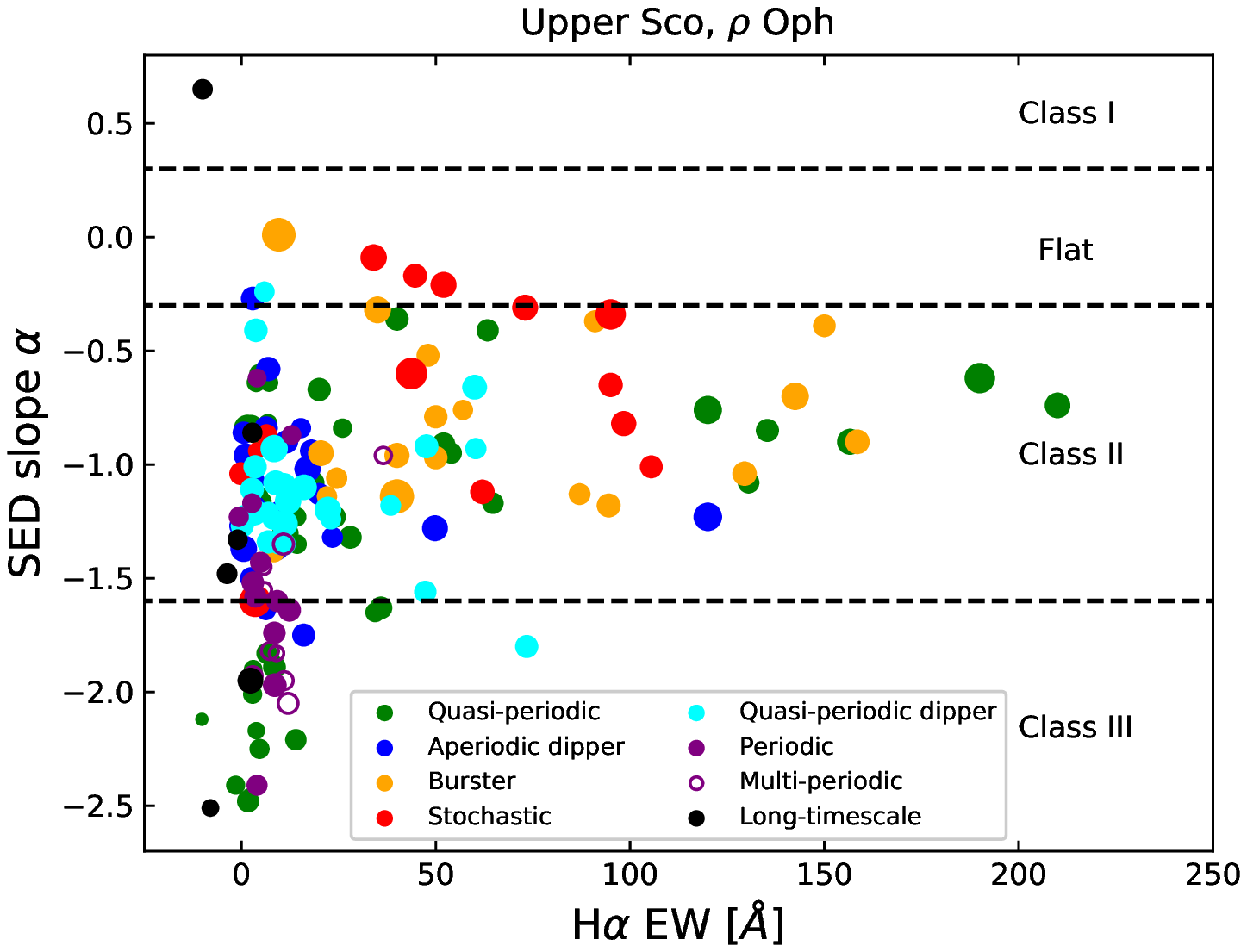}
\caption{SED slope ($\alpha$) versus H$\alpha$ equivalent width for Taurus and Upper Sco/$\rho$~Oph. These quantities are measures of infrared excess and accretion strength, respectively. We use colors to denote different variability types and search for correlations between flux behavior and inner disk properties. The most prominent patterns identified include consistently low gas emission for the dippers and periodic stars, as well as very weak disks (i.e., low infrared excess) for the latter. As in previous figures, points are scaled in size according to amplitude to the 1/3 power. 
}
\label{Halpha_alpha}
\end{figure}

In Figure~\ref{Halpha_alpha} we show the photometric variability classes in a disk diagnostics plot, with $\alpha$ measuring infrared excess strength (and consequently disk geometry and accretion) versus equivalent width of H$\alpha$ emission (where available in the literature), measuring gas accretion and outflow strength. The equivalent widths were adopted from the literature\footnote{\citet{1988cels.book.....H, 1993PASP..105..686B, 1998AJ....115.2491K, 2000AJ....119.1881A, 2000AJ....120.2114M,  WhiteGhez2001, JensenAkeson2003, 2003ApJ...582.1109W, 2003ApJ...592..266M, 2004ApJ...616..998W, Muzerolle+2005, Gudel+2007, Wahhaj+2010, Duchene+2017, Kraus+2017}.}, and are therefore non-simultaneous with the {\em K2} variability assessed here. 
Besides their photometric variability, the variability of H$\alpha$ line profiles and line strengths  is well documented
\citep[e.g.][]{zsidi2022} in accreting T Tauri stars. 
Despite being prone to scatter, Figure~\ref{Halpha_alpha} does show a few notable trends.

Stars with periodic signals are segregated towards lower infrared excess and smaller gas emission, as expected if we have a clear view towards the stellar photosphere, unaffected by line-of-sight dust or gas. Stars with quasi-periodic or aperiodic dipping light curves also tend to have smaller gas emission, though span a range of infrared excess. The quasi-periodic symmetric stars typically have somewhat larger gas emission and also span a range of infrared excess.  There are few bursters, but they tend towards Class II type disks, with a range of gas emission strengths.

For comparison, we also show the same diagram for the \cite{cody2018} sample of Upper Sco and $\rho$ Oph stars.  The larger population there exhibits similar behavior with respect to the light curve categories, though the trends described above are not quite as clear, with more overall spread.  There is a well-populated concentration of quasi-periodic dippers at moderate infrared excess and low gas emission. We note that using infrared colors instead of the SED slope $\alpha$ typically hides the trends discussed above.

The observed variability may not only be tied to disk emission and gas accretion strength, but also to viewing geometry. It has long been hypothesized \citep{bouvier1999} that dipper stars must be viewed at high inclinations in order for dusty material in the inner disk to block our line of sight. This idea was recently questioned by \cite{ansdell2020}, who noted several dippers with nearly face-on measured disk inclinations. We further assess the correlation between inclination and variability type here.

\begin{figure}
\epsscale{1.2}
\plotone{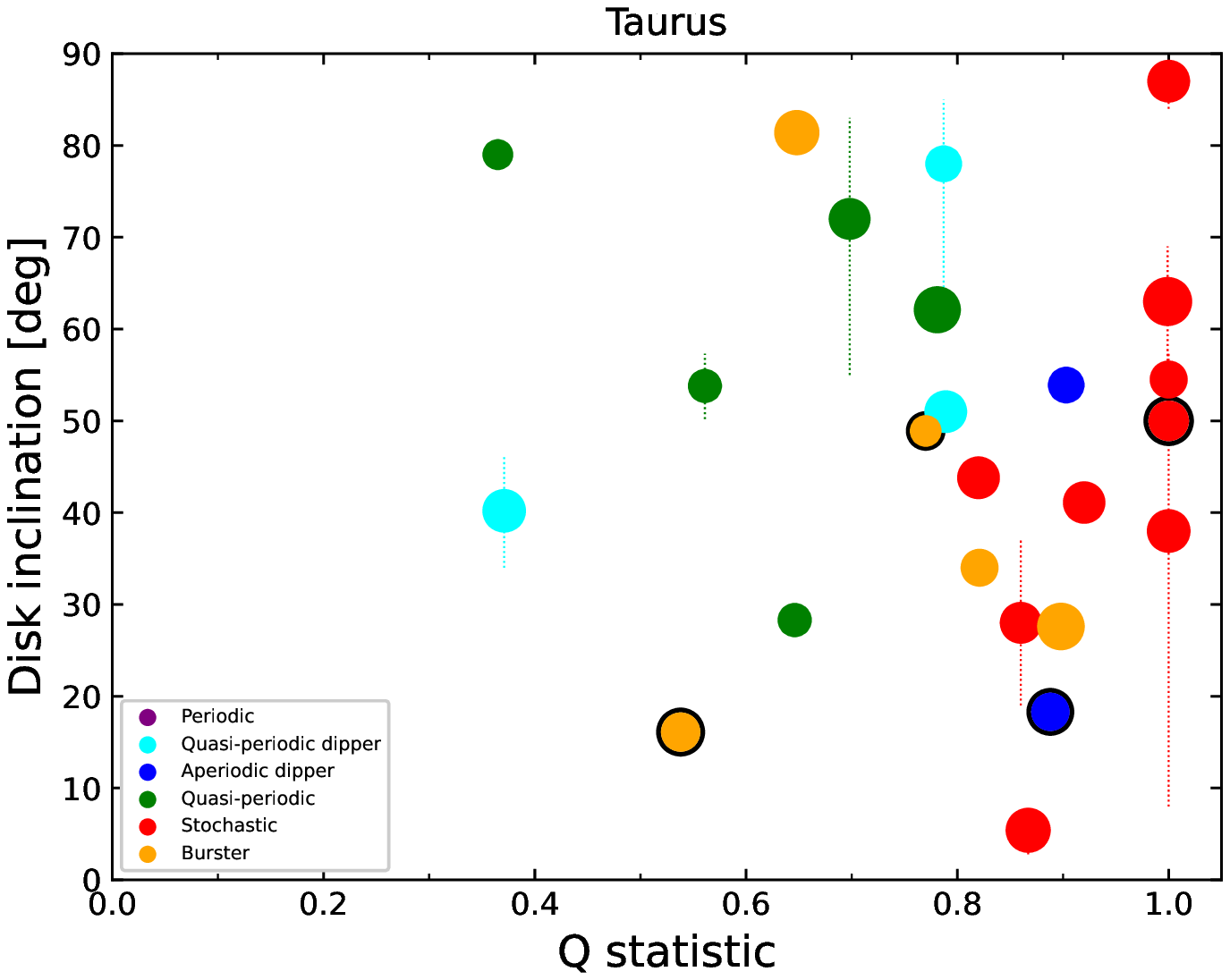}
\plotone{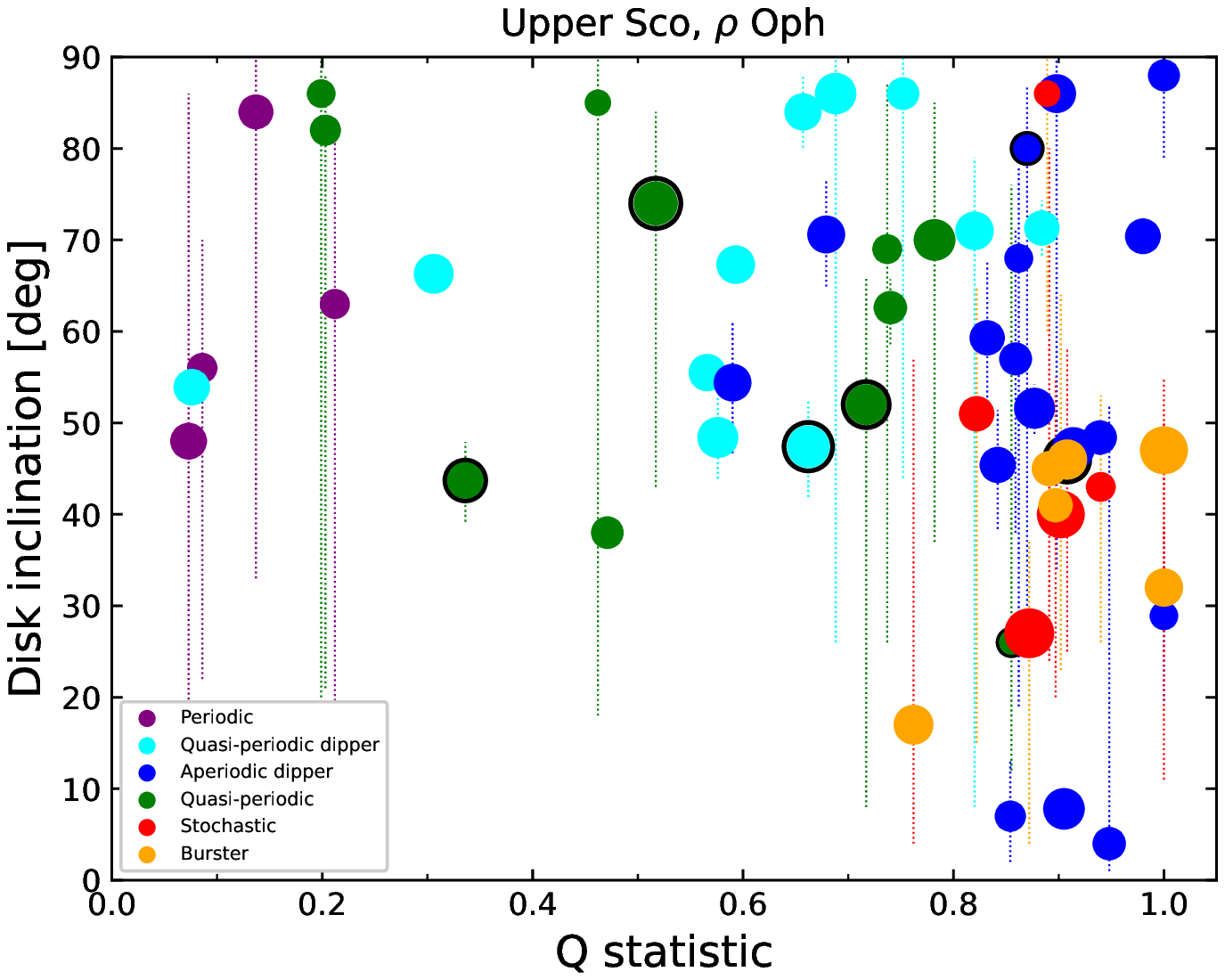}
\caption{Available inclination measurements versus $Q$ value, and colored by variability class. Points are scaled in size according by variability amplitude to the 1/3 power. Those circled in black denote blends (e.g., multiple star systems). These data, particularly for Upper Scorpius and $\rho$~Oph (bottom panel) suggest that single (non-binary) periodic and quasi-periodic sources are associated with disk inclinations of at least 30-40\arcdeg.}
\label{taurusinclinations}
\end{figure}

To do so, we assemble an updated inclination dataset from millimeter and submillimeter observations of Taurus disks.  Recent high-quality disk inclinations are available for a decent fraction ($\sim$25\%) of our sample \citep{aizawa2020, long2019, tripathi2017, schaefer2009, villenave2020, ansdell2020}. In Figure~\ref{taurusinclinations}, we show the inclinations as a function of our calculated $Q$. Of note  is that there is no inclination data for the periodic sources in our sample, consistent with our finding above that the periodic sources have small infrared excesses; the disks surrounding these sources are apparently weak in the millimeter and submillimeter as well, according to their infrared spectral energy distributions. 

Figure~\ref{taurusinclinations} also compares the Taurus inclinations and $Q$ values to those derived for $\rho$~Oph and Upper Sco.  For the latter, the inclinations are from \cite{barenfeld2017} and \cite{ansdell2016}; $Q$ data comes from \cite{cody2018}. All values are color coded by light curve morphology type, to enable exploration of the variability properties as a function of inclination. We circle in black points for which multiple sources are blended into the light curve; in these cases, it is difficult to uniquely associate light curve behavior with geometry. For several binary systems, including DK Tau AB (EPIC 210777988), HK Tau AB (EPIC 247799571), and UZ Tau AB (EPIC 248009353), there are two very different measured disk inclinations, and we omit them from the Taurus panel. 

Our main conclusion from the inclination-$Q$ comparison is that periodic and quasi-periodic sources are almost exclusively found at moderate to high inclination (40--90\arcdeg), while the disk inclination distribution of aperiodic sources appears largely isotropic. This finding is particularly significant for the dipper stars, which split into quasi-periodic and aperiodic subtypes. The differing inclination distributions for these subtypes suggests that they may originate from distinct phenomena, or different locations, in terms of disk radius. 

\subsection{Variability and Stellar Properties}

\begin{figure}
\epsscale{1.25}
\plotone{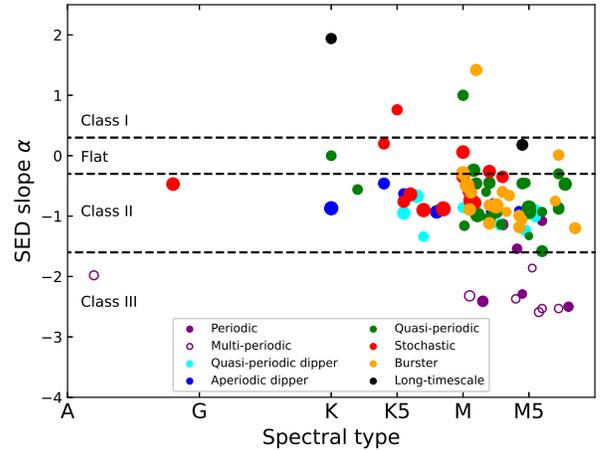}
\caption{Infrared excess slope as a function of spectral type (a proxy for mass) and variability type. We use this diagram to assess correlations in variability type as a function of stellar mass and disk evolutionary status, but do not detect appreciable
trends apart from periodic sources appearing almost exclusively in the class III set.}
\label{alphaspt}
\end{figure}

In Figure~\ref{alphaspt} we show the distribution of $\alpha$ measuring infrared excess slope with spectral type, again color coded by photometric variability type. There are no clear trends in variability type with spectral type, a proxy for stellar mass in the early pre-main sequence phase.  An alternate plot using color instead of spectral type (not shown) also reveals no trends. 

We also do not see any tendency toward lower amplitude at higher mass and temperature, which was a tentative trend in the Upper Sco dataset \citep{cody2018} as well as a finding of \citet{venuti2021}. However, there are only a few stars in our Taurus sample earlier than spectral type K, so this is a weak conclusion. What we do see is that the collection of low-amplitude ($<1\%)$ {\em periodic} variables is clustered at low mass (e.g, purple points of spectral type M in Figure~\ref{alphaspt}). This may again be result of the lack of early type stars in the sample, as well as the younger age of Taurus, as compared to Upper Sco. 

\section{Comparisons with other {\em K2} Clusters}

With {\em K2} photometry available now for more than one star-forming region dominated by low-mass YSOs, we are in a position to perform a comparison of variability properties with age. Several Figures (\ref{qmplot}, \ref{amptimescale}, \ref{Halpha_alpha}, \ref{taurusinclinations} ) 

already compare Taurus, at 1--5~Myr of age, with $\rho$~Oph ($\sim$1~Myr) and Upper Sco ($\sim$5--10~Myr). We find several significant differences between these clusters. 

First, there is a a significantly lower fraction of low-$Q$ ($<$0.4) systems in Taurus, as seen in Figure~\ref{qmplot}. There is a conspicuous gap in the $Q$-$M$ diagram from $Q\sim$0.2-0.35. This gap is difficult to explain as an evolutionary effect since it is not present in either the younger $\rho$~Oph dataset or the older Upper Sco dataset. 

In addition, the measured incidence of quasi-periodic dippers (QPDs) in Taurus is only about half that of $\rho$~Oph and Upper Sco. The errorbars are substantial, though, and it is possible that the younger two regions (Taurus and $\rho$~Oph) both have a lower QPD rate of roughly 11\%, while Upper Sco remains elevated at at least 16\%. This trend of increasing dipper fraction at older age would be in line with the tendency for dipper behavior to occur for stars with weaker inner disks, as indicated by smaller near and mid-infrared colors \citep{cody2018}. 

Turning to the top portion of the $Q$-$M$ diagram in Figure~\ref{qmplot}, we also see more quasi-periodic bursters (orange points with $Q<0.85$) in Taurus as compared to $\rho$~Oph and Upper Sco, but somewhat less extreme flux asymmetries (lower $M$) overall. 

There is also a trend in overall amplitudes decreasing as a function of age. Whereas some of the $\rho$~Oph burster light curves vary by a factor of almost three (0.5 in log space, as seen in Figure~\ref{amptimescale}), the amplitudes in Taurus and Upper Sco rarely exceed 100\%. Furthermore, the bottom envelope of the amplitude distribution decreases with age. Quasi-periodic targets in Upper Sco extend down to amplitudes as small as 0.001, whereas they only reach $\sim$0.005--0.01 in Taurus. In $\rho$~Oph, we do not see any amplitudes smaller than 0.025.

Timescales of variability are presumably dominated by several phenomena, including accretion fluctuations and stellar rotation or inner disk orbits (at a similar timescale if at co-rotation). Among the three regions $\rho$~Oph, Taurus, and Upper Sco, we do not observe a significant difference in the overall distribution of timescales (Figure~\ref{amptimescale}). When broken down into individual variability types, we only detect a difference among quasi-periodic dippers. The timescales for the QPDs extend to somewhat larger values for the older population of Upper~Sco than they do in $\rho$~Oph and Taurus. The origin of this effect is unclear but may be connected with the disk clearing that occurs at older ages.

We also detect a stronger tendency toward periodic bursting behavior in the younger $\rho$~Oph and Taurus populations, as compared to Upper Sco. This is evident in the rightward shift of the orange points (bursters) from the top $Q$-$M$ diagram in Figure~\ref{qmplot} to the bottom one. In Taurus, the behavior at  $Q>0.5$ is mostly low-M, i.e., burst-dominant, whereas in Upper Sco we see a wide range of light curve asymmetry here.  

All of the above trends are consistent with the variability patterns in younger populations being more dominated by the accretion properties of the YSOs, than either the properties of the stars (which are the origin of the periodic light curve signatures indicating stellar rotation) or the properties of the optically thin layers of the disks (which are the likely origin of the dipper light curve effect). The explanation for this is that the Taurus stars have — on average — less evolved disks than those in Upper Sco. In Upper Sco, if the accretion rates are somewhat lower and the disks somewhat more evolved, variability could be dominated by the kinematics of the rotating dust near the co-rotation radius. Our observations support the idea that the accretion timescale is shorter than that of inner disk dispersal, as measured by, e.g., near-infrared excess.

\section{Discussion and Summary}

Young stars of the Taurus Molecular Clouds are the fourth population of $<10$ Myr old stars for which high-precision, high-cadence optical photometric monitoring data are available from space-based platforms. We have presented here a study of the young disk-bearing stars in Taurus observed by K2, and applied quantitative metrics to classify their light curve behavior. 

Within Taurus, we found that {\em all} of the disk-bearing stars display variability detectable above the photometric noise floor. Using the $Q$ and $M$ statistics developed by \citet{cody2014}, we have divided the light curves into morphology categories, as listed in Table~3. In Taurus, the dominant mode of variability is ``quasi-periodic symmetric" behavior, in which a pattern that favors neither flux increases (bursts) nor decreases (dips) emerges. The second most common variability type is bursting, the majority of which in Taurus is quasi-periodic. Dipper behavior is less common in Taurus than its incidence in the other clusters. More rare variability types include strictly periodic stars and those with long-timescale trends. In general, around 70\% of the light curves have repeating patterns that we label as periodic or quasi-periodic, while the remaining 30\% display more erratic, and typically higher amplitude, variability. These estimates are based on the $Q$ values listed in Table~2 and displayed in Figure~\ref{qmplot}. 

We have previously discussed the origins of these different variability morphology types, and our findings in Taurus largely support a consistent picture. Several different driving mechanisms are at work, with the possibility for multiple to operate at the same time in individual stars. 

The simplest variability categories to explain are the periodic and multi-periodic types. We have shown that stars with perfectly repeating (and often sinusoidal) patterns are most often class III sources with very weak disks. Thus we are likely seeing the rotational modulation of one or more magnetic starspots, since there is little accretion or disk material to introduce other genres of flux variation. 

The bursters, on the other hand, likely represent cases for which accretion flow variations and associated stellar surface hot shock regions dominate the flux changes. \cite{ribas2014} and \cite{fedele2010} showed that accretion rates decrease over time, and hence we would expect to see bursting behavior tapering off as stars evolve. 
We have detected a decrease in the {\em quasi-periodic} burster incidence from $\rho$~Oph through Taurus to older Upper Sco that may reflect this trend. The Upper Sco sample included a population of aperiodic bursters, which could point to a shift from rotationally modulated hotspots to more erratic gas flow as YSOs age.

Long-timescale variability appears to be almost exclusively associated with class I or flat spectral energy distributions, and hence more embedded sources. We are likely observing flux changes connected with copious circumstellar material and light reflected off of it that originates on or near the stellar surface. There may be many more of these long timescale variables, but higher levels of extinction precludes good photometric detection and precision with {\em Kepler}. 

The physical driver of quasi-periodic variations remains difficult to pinpoint, as it may be a combination of spot modulation (either cool magnetic spots or hot accretion spots) and obscuration by orbiting circumstellar material near co-rotation. Future multi-color photometric monitoring should help distinguish the different mechanisms here. 

Finally, dipper variability has typically been ascribed to occultation of the central star by elevated dusty material at or near the inner disk edge. Our data support this hypothesis but suggest that there may be different origins for quasi-periodic and aperiodic dipper behavior. This idea comes from the finding that single (i.e., non-binary), quasi-periodic dipper stars in the $\rho$~Oph, Taurus, and Upper Sco samples all have measured disk inclinations of at least 40\arcdeg\ (at least for the fraction of YSOs where interferometric data is available). The set of aperiodic dippers, on the other hand, seems to have a relatively isotropic distribution of inclinations, with some targets hosting nearly face-on disks.  

Few theoretical models have endeavored to explain the full suite of YSO variability elucidated in the recent continuous, high-cadence space telescope data. However,  \cite{robinson2021} demonstrated that they can populate much of the parameter space
at $M < 0.25$ (symmetric to bursting light curves) and $Q < 0.9$ (not completely stochastic light curves) with magnetospheric accretion disk models of varying co-rotation radii; additional important parameters include the detailed configuration of the magnetic field, and the fractional area of the accretion flow on the surface of the star.  
The models are somewhat idealized in their geometry and spot parameters, but nevertheless provide valuable insight.
Specifically, \cite{robinson2021} create synthetic lightcurves for systems spanning appropriate ranges in physical and geometric parameters describing the infall of material from an inner disk region onto a central star. They then analyze noised-up versions of the lightcurves using the $Q$ and $M$ metrics. A major determinant of the lightcurve behavior is source inclination, with \cite{robinson2021} finding that the high-inclination sources have purely periodic low $Q$ values, the intermediate inclination have quasi-periodic, intermediate $Q$ values, and the low inclination have stochastic, high $Q$ values, where we essentially are viewing the stochastic infall along the field lines directly, unmodulated by stellar rotation. However, these are stated as ``typical" trends, with the $Q$ values for any individual source also influenced by other parameters in the model.  The $M$ metric was stated as most influenced by parameters associated with the magnetic field strength and geometry, along with source inclination.

At this juncture, it is possible that not all of the observables required to explain different variability modes are amenable to observational constraint. For example, magnetic field strength and geometry is only available for a handful or bright targets. Through our work on YSOs observed during the {\em K2} Mission, we have confirmed that no single stellar or circumstellar parameter determines the time domain flux properties. The light curve morphology, timescale and amplitude are shaped by disk evolutionary state (and hence stellar age), system inclination, accretion rate, and likely magnetic properties as well. Further monitoring of both low-mass and intermediate-mass YSOs by the {\em TESS} satellite may soon help further untangle the physical drivers of variability, enabling renewed modeling efforts and constraints on the nature of the inner disk region where terrestrial planet formation may be ongoing at this epoch.

\begin{acknowledgements}
We thank the referee for constructive comments. This paper includes data collected by the {\em K2} mission, for which funding was provided by the NASA Science Mission directorate. This work was additionally supported by {\em K2} Guest Observer program GO13117 and NASA award 80NSSC17K0026.
\end{acknowledgements}

\bibliography{K2Taurus.bib}

\clearpage
\appendix

\section{Lightcurves for Individual Sources}

In this Appendix we show light curves for disk-bearing (Figure~\ref{alllcs}) and non-disk-bearing (Figure~\ref{allmaybelcs}) sources towards Taurus.

\begin{figure*}[!b]
\epsscale{0.9}
\plotone{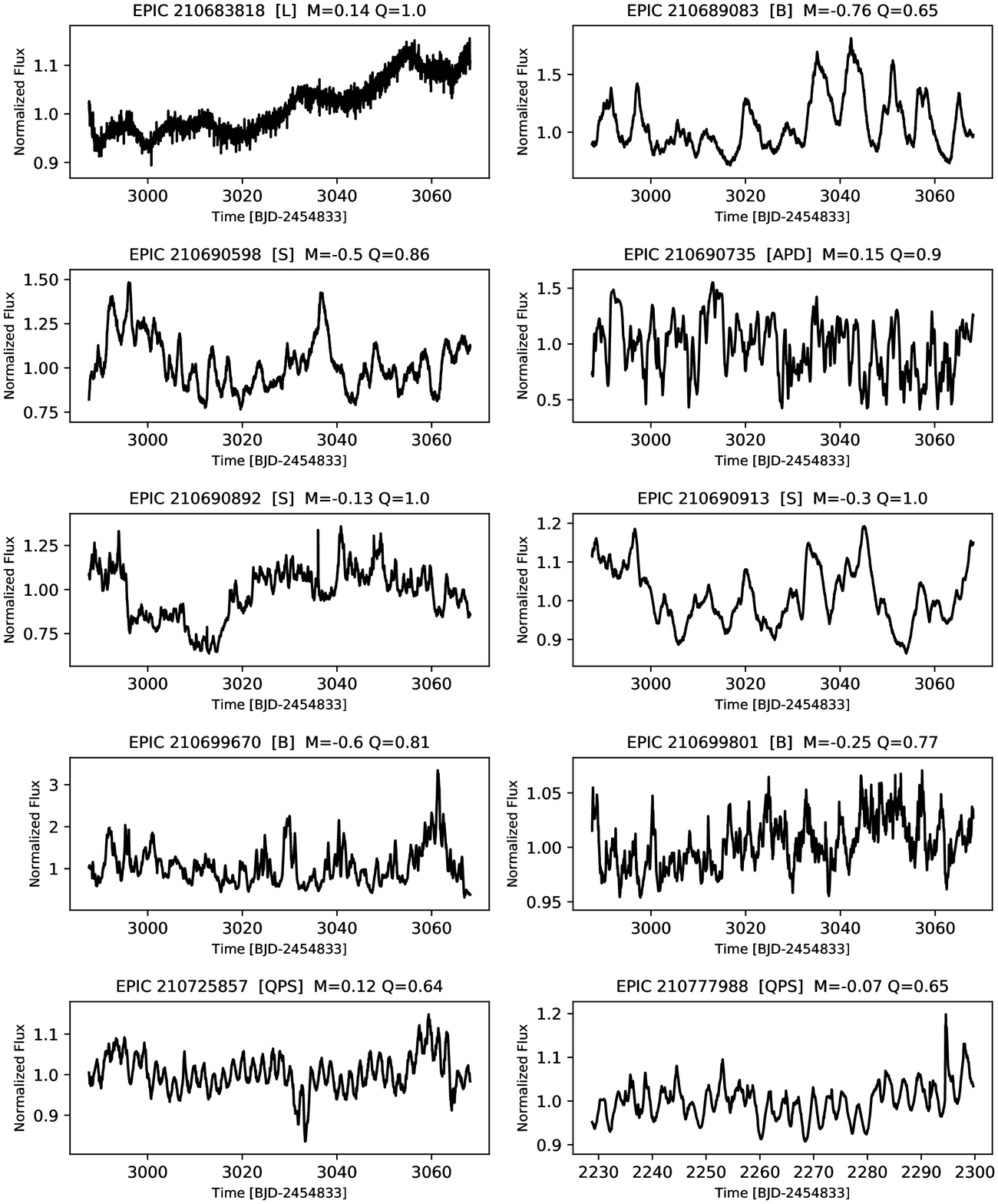}
\caption{Median-normalized light curves of disk-bearing stars in Taurus observed with {\em K2}, in order of EPIC identifier. Figure labels include the variability type from Table~2, namely "P" = strictly periodic behavior, "MP" = multiple distinct periods, "QPD" = quasi-periodic dippers, "QPS" = quasi-periodic symmetric, "APD" = aperiodic dippers, "B" = bursters, "S" = stochastic stars, "L" = long-timescale behavior that doesn't fall into the other categories.
Values of flux symmetry metric $M$ and the quasi-periodicity metric $Q$ from Table~2 are also provided.}
\label{alllcs}
\end{figure*}

\clearpage

\addtocounter{figure}{-1}
\begin{figure*}
\epsscale{0.90}
\plotone{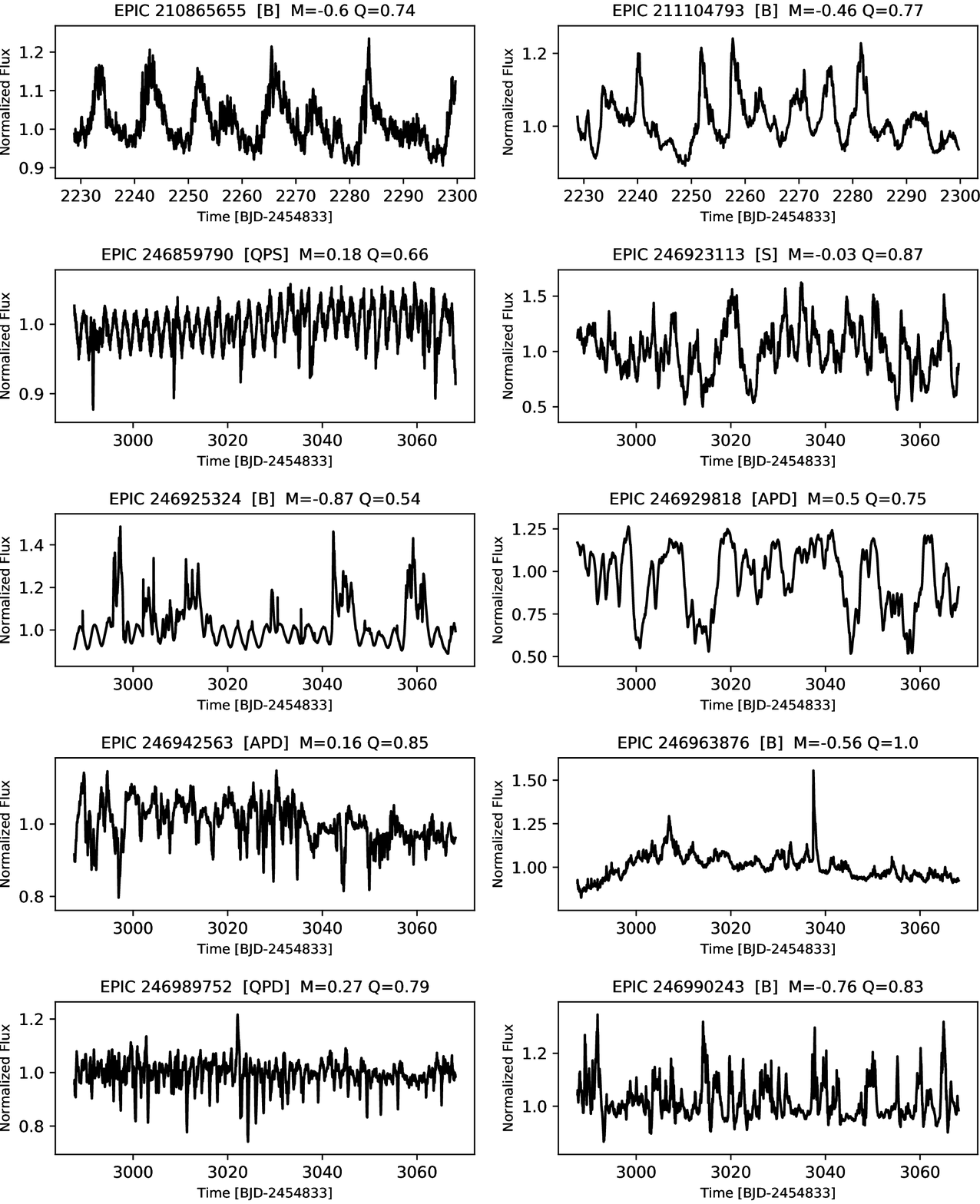}
\caption{Cont.}
\end{figure*}

\clearpage

\addtocounter{figure}{-1}
\begin{figure*}
\epsscale{0.90}
\plotone{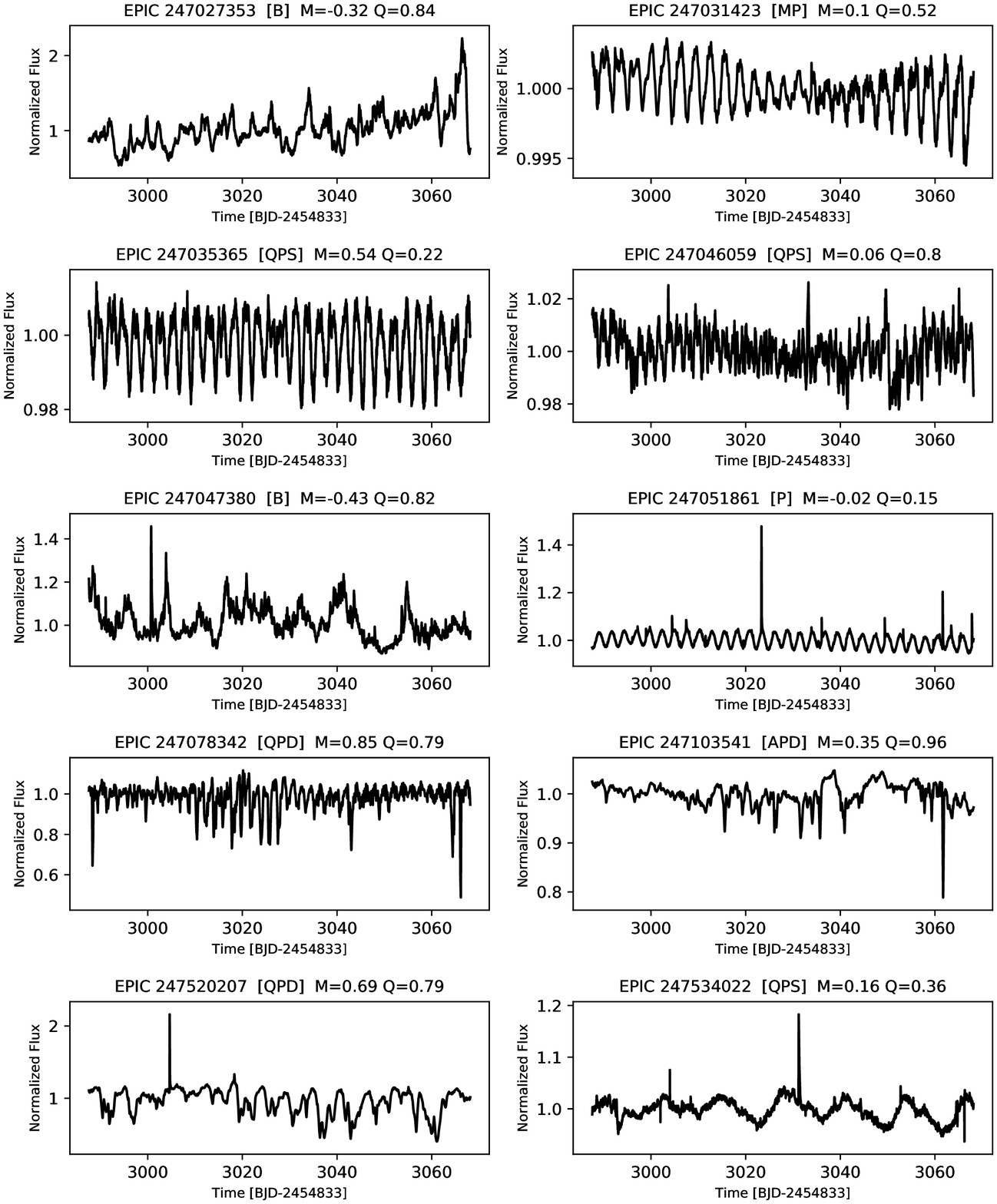}
\caption{Cont.}
\end{figure*}

\clearpage

\addtocounter{figure}{-1}
\begin{figure*}
\epsscale{0.90}
\plotone{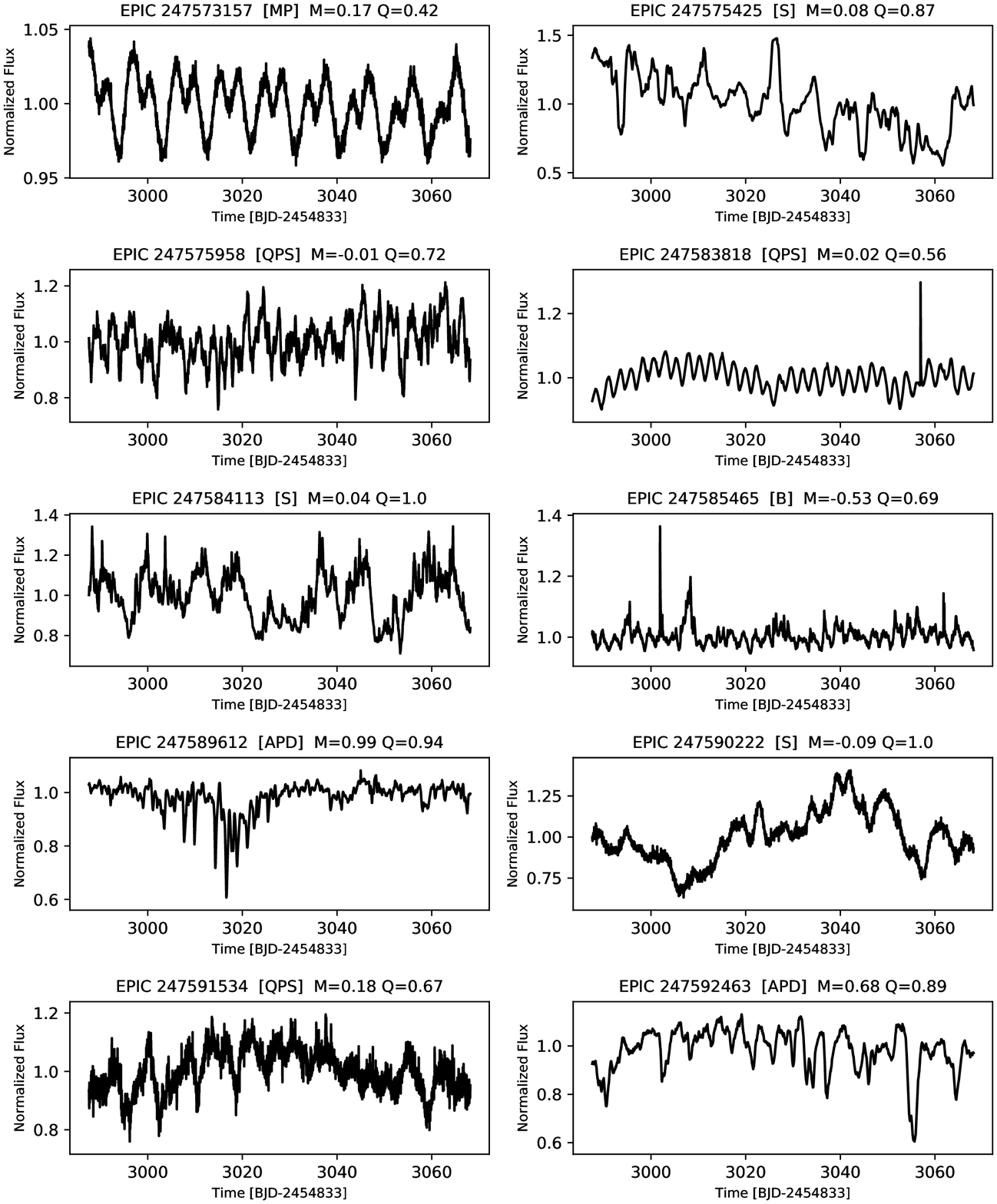}
\caption{Cont.}
\end{figure*}

\clearpage

\addtocounter{figure}{-1}
\begin{figure*}
\epsscale{0.90}
\plotone{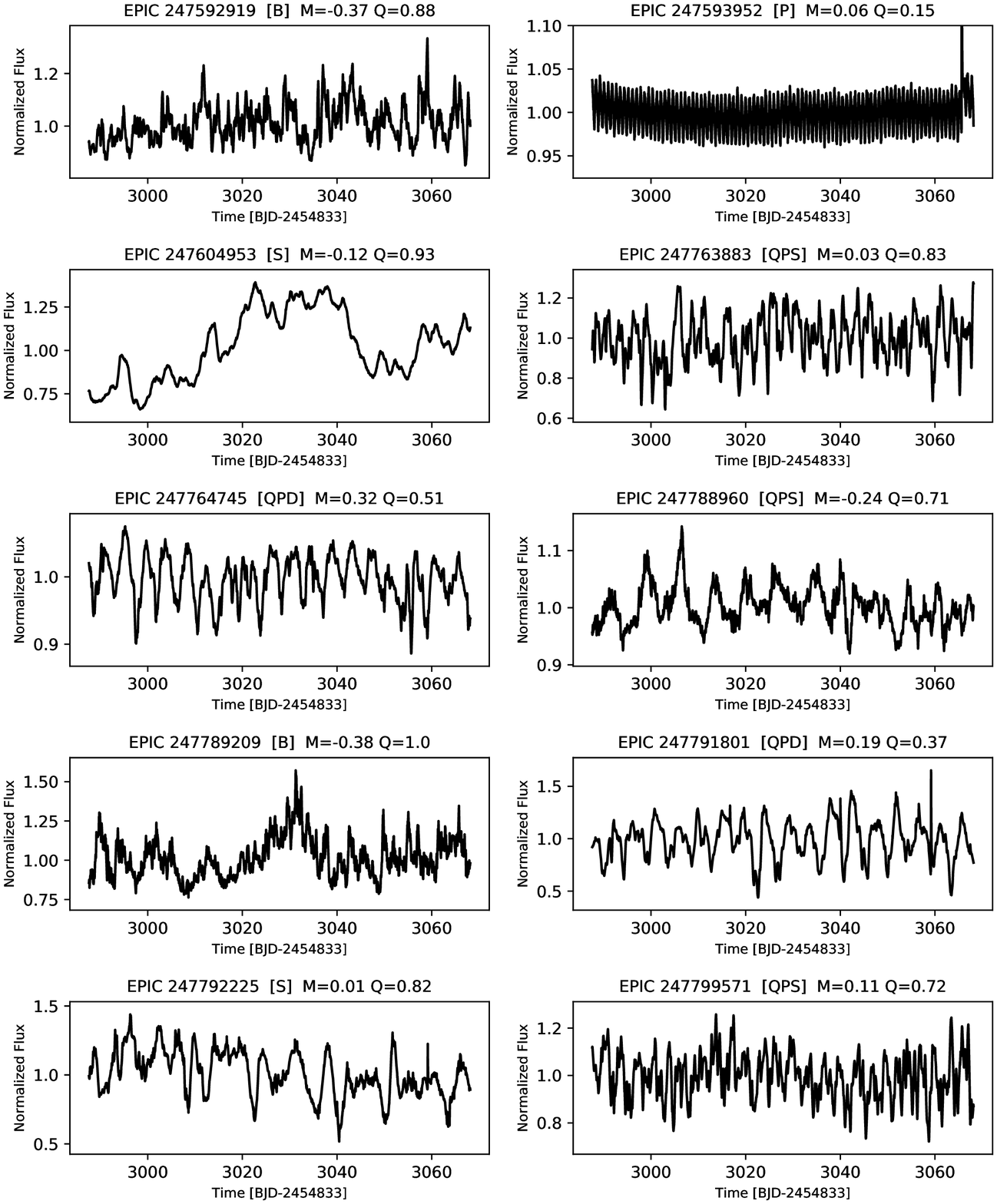}
\caption{Cont.}
\end{figure*}

\clearpage

\addtocounter{figure}{-1}
\begin{figure*}
\epsscale{0.90}
\plotone{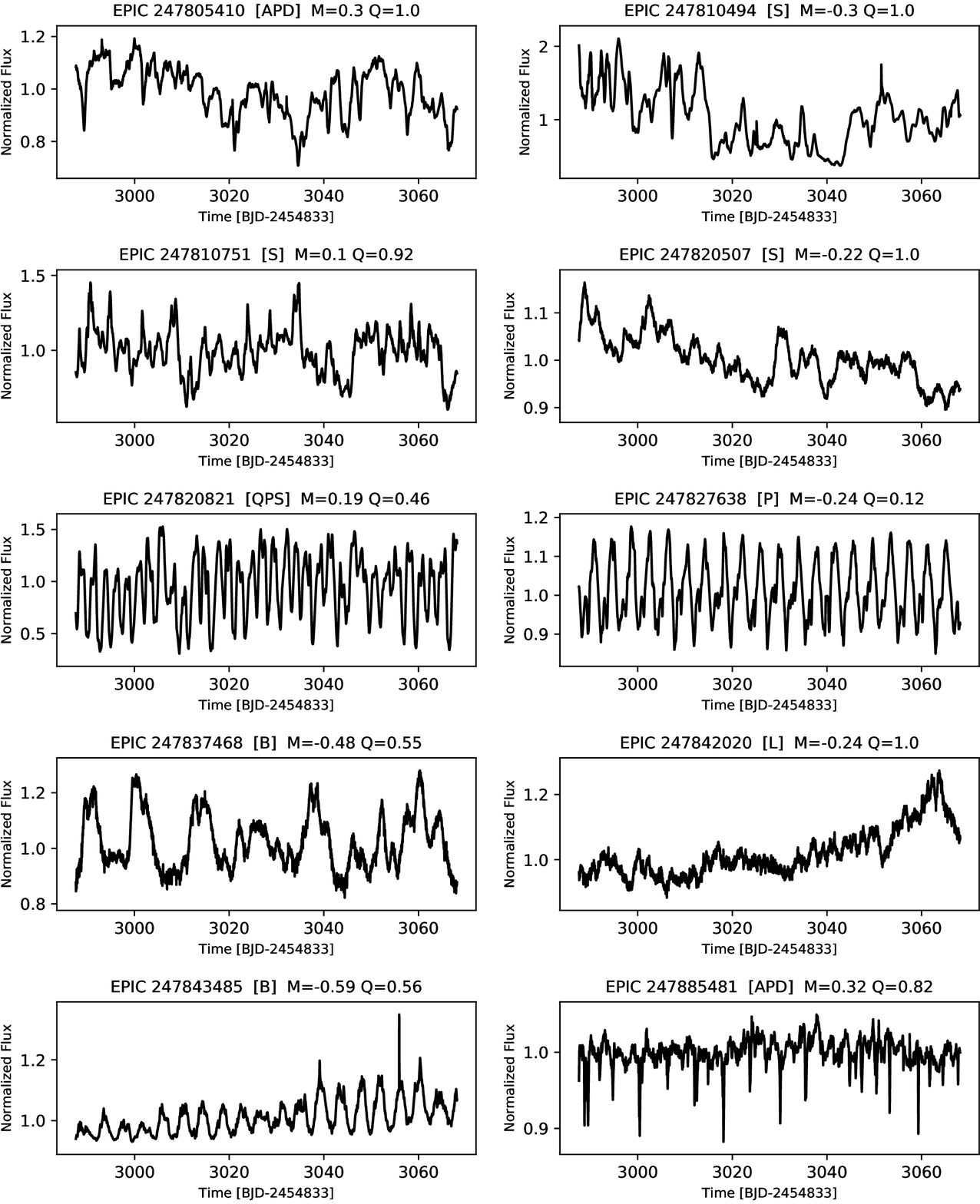}
\caption{Cont.}
\end{figure*}

\clearpage

\addtocounter{figure}{-1}
\begin{figure*}
\epsscale{0.90}
\plotone{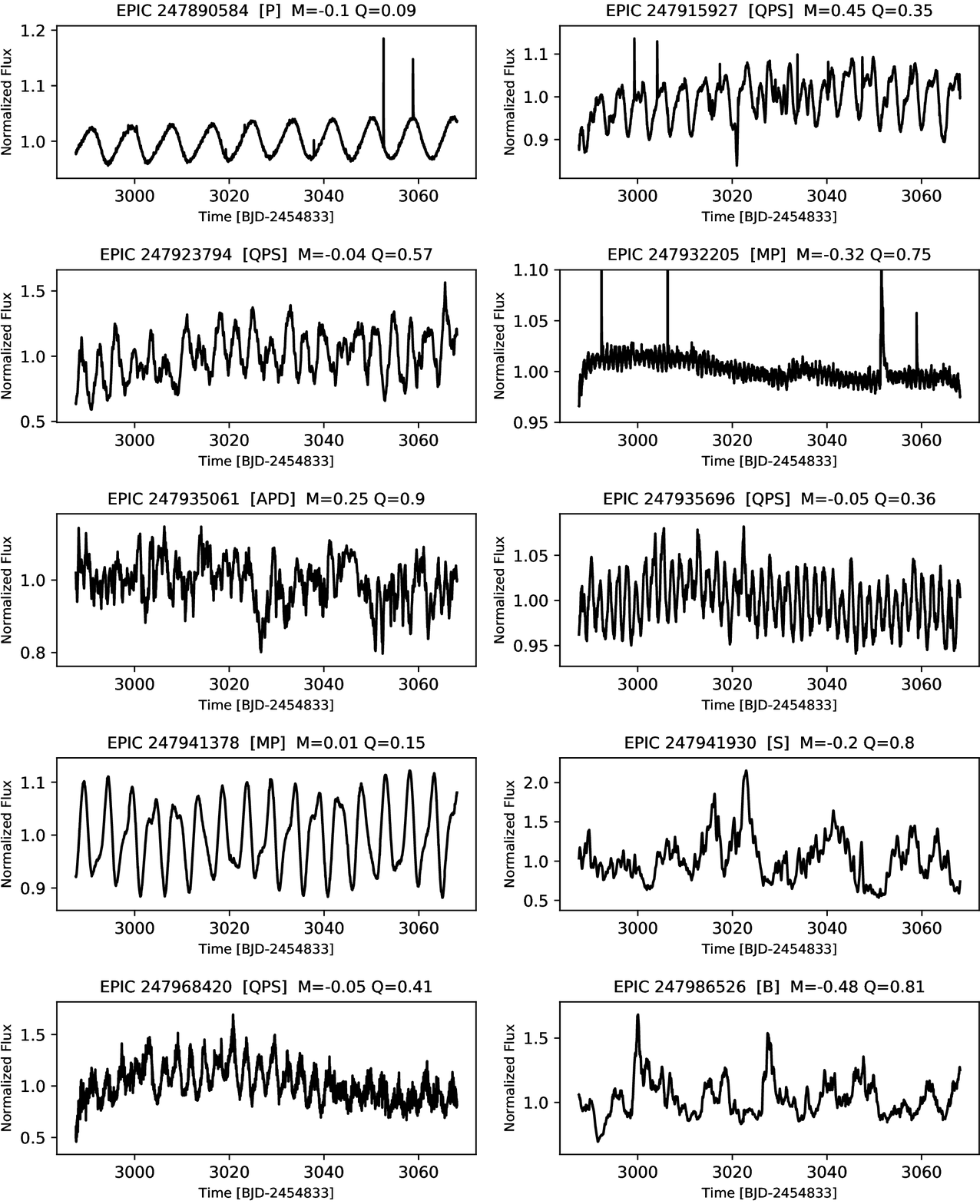}
\caption{Cont.}
\end{figure*}

\clearpage

\addtocounter{figure}{-1}
\begin{figure*}
\epsscale{0.90}
\plotone{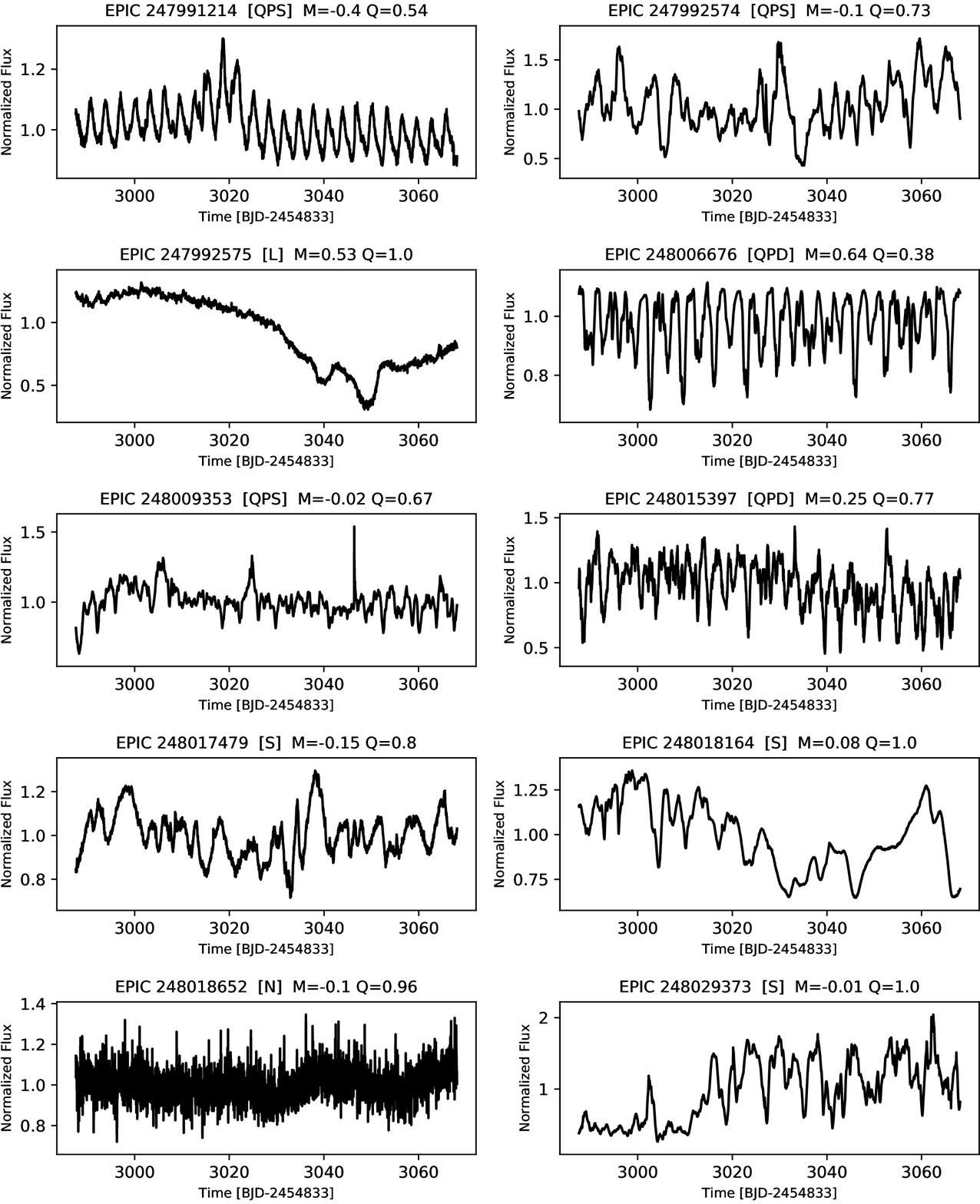}
\caption{Cont.}
\end{figure*}

\clearpage

\addtocounter{figure}{-1}
\begin{figure*}
\epsscale{0.90}
\plotone{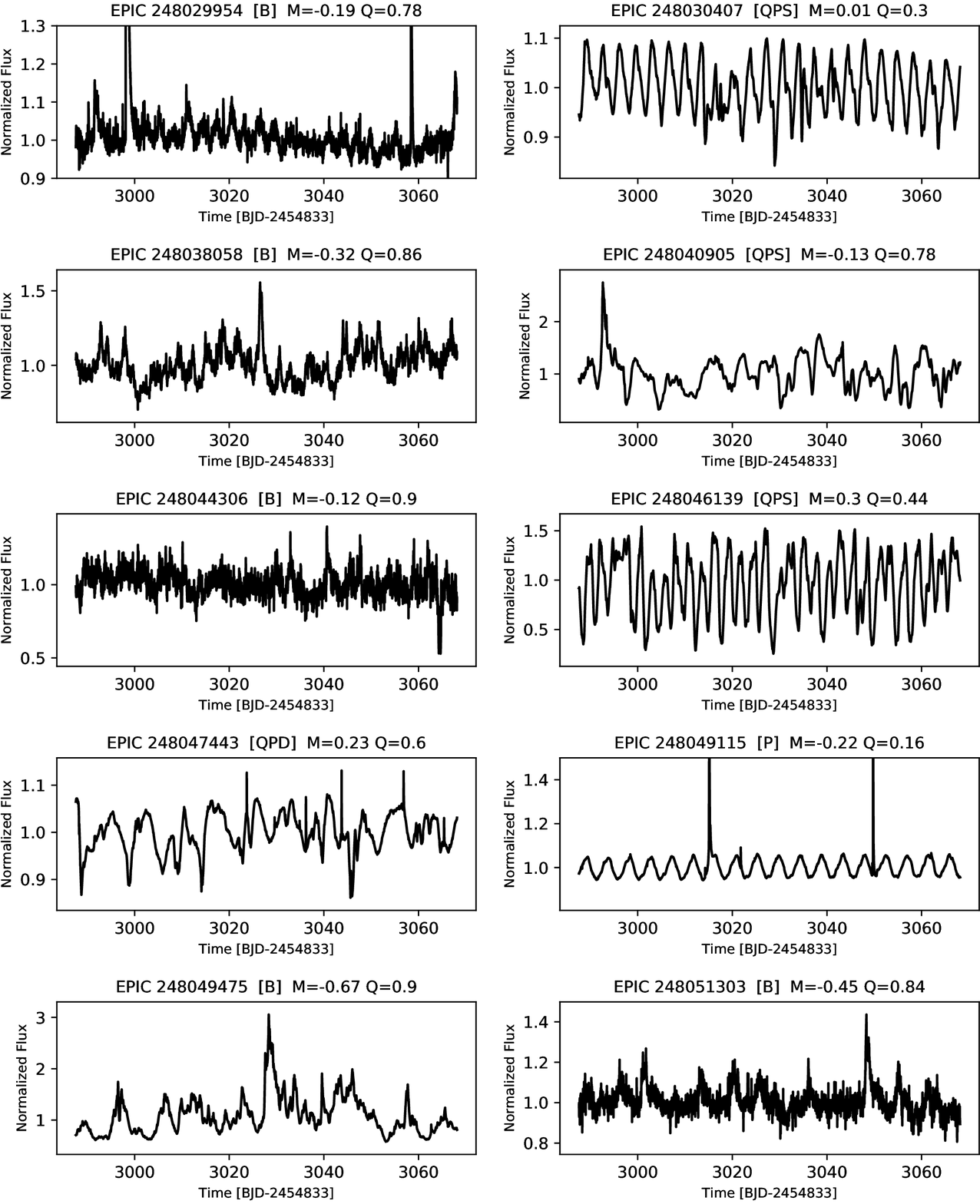}
\caption{Cont.}
\end{figure*}

\clearpage

\addtocounter{figure}{-1}
\begin{figure*}
\epsscale{0.90}
\plotone{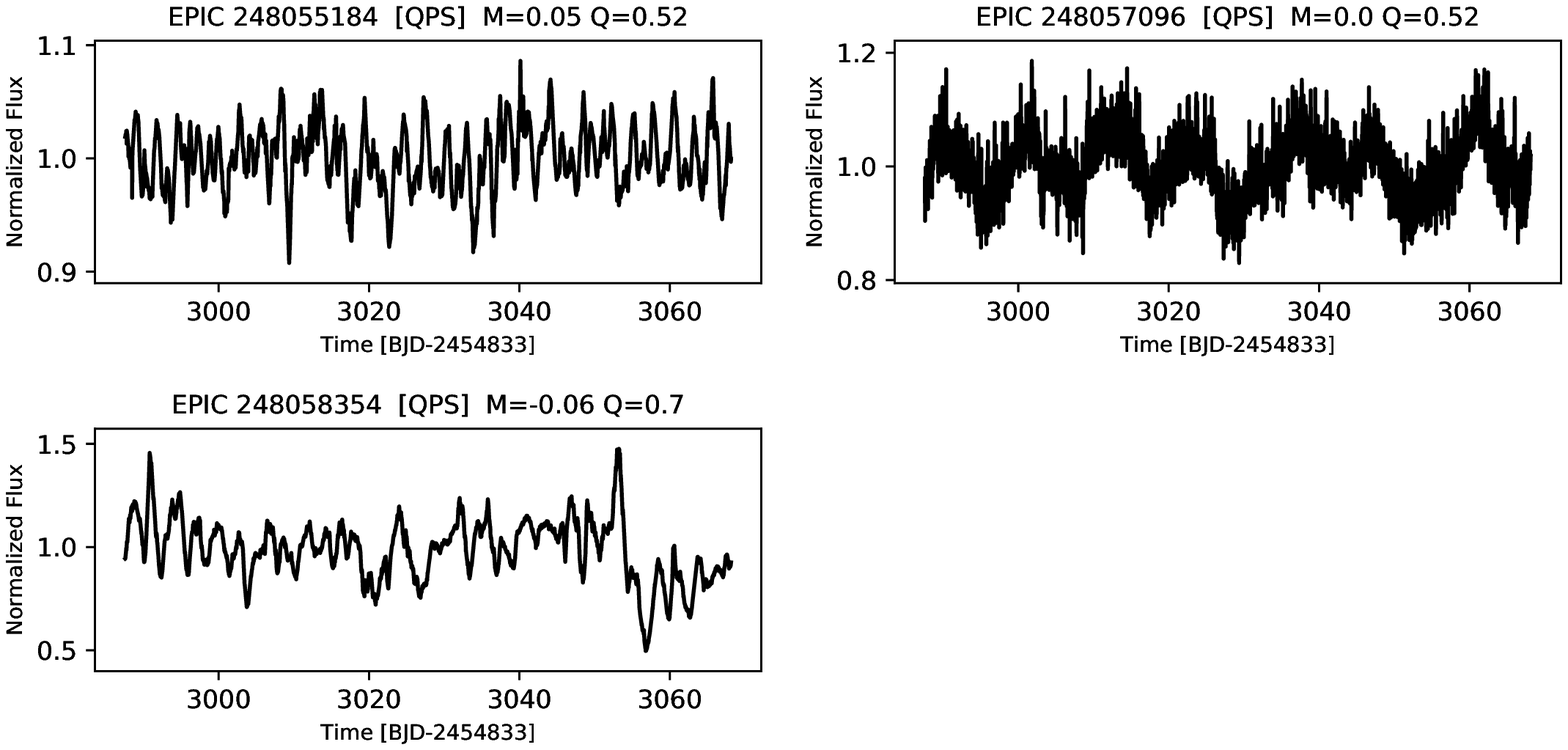}
\caption{Cont.}
\end{figure*}

\begin{figure*}
\epsscale{0.90}
\plotone{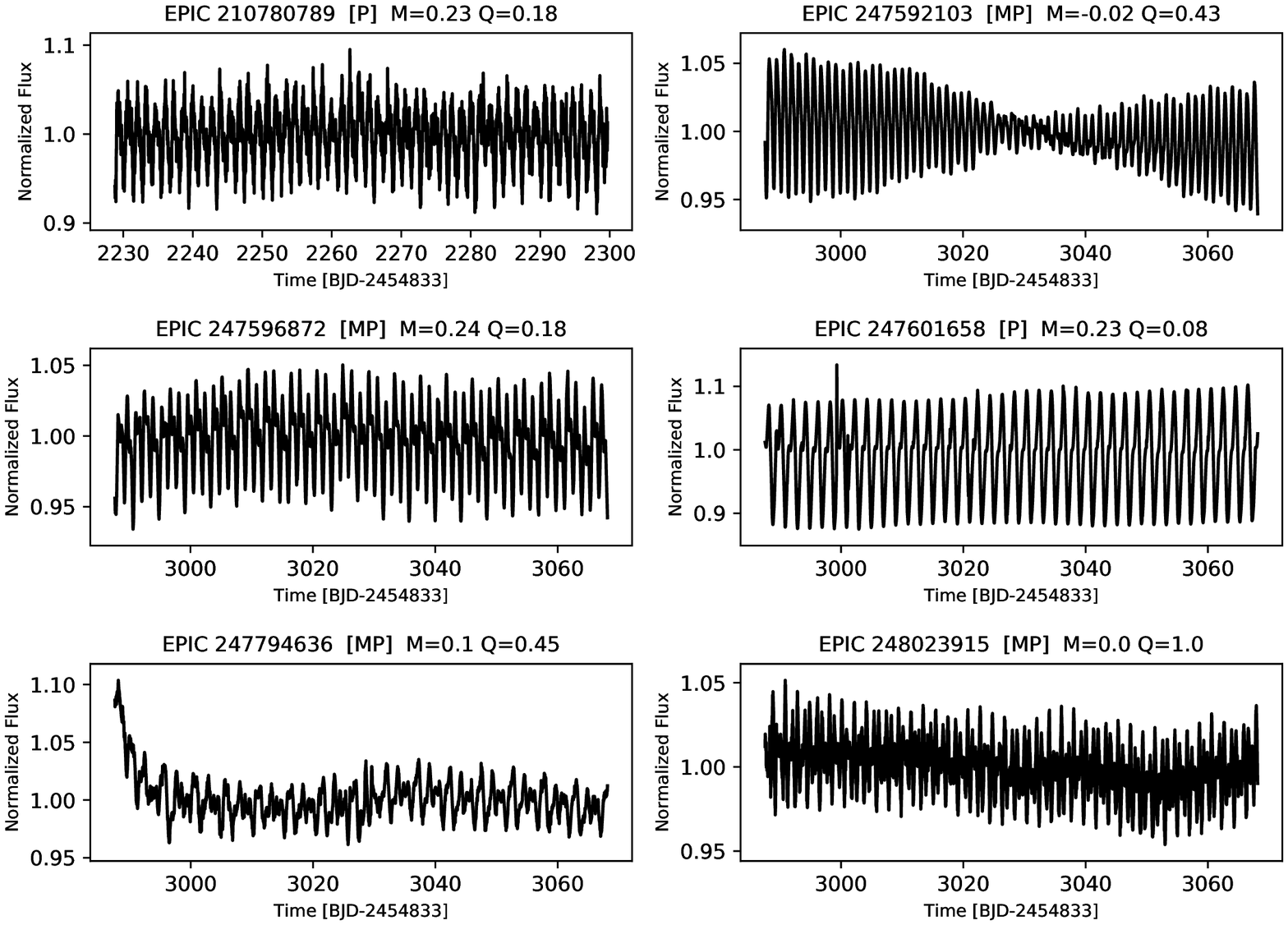}
\caption{Light curves of {\em candidate} disk-bearing stars in Taurus observed with {\em K2}, in order of EPIC identifier.  Figure labels include the variability type from Table~2, namely
"P" = strictly periodic behavior, "MP" = multiple distinct periods, "QPD" = quasi-periodic dippers, "QPS" = quasi-periodic symmetric, "APD" = aperiodic dippers, "B" = bursters, "S" = stochastic stars, "L" = long-timescale behavior that doesn't fall into the other categories.
Values of flux symmetry metric $M$ and the quasi-periodicity metric $Q$ from Tabler~2 are also provided.}
\label{allmaybelcs}
\end{figure*}
 
\clearpage
\section{Comparison with R20}

As a check on the timescales derived for our periodic and quasi-periodic sources (e.g., those with $Q<0.85$), we performed a comparison with the results of R20 who studied periodic sources in Taurus.  Although ostensibly the same data set was used in the two studies, there are methodological differences, as we have used a combination of Fourier transforms and autocorrelations instead of Lomb Scargle periodograms. In addition, the period searched light curves adopted in R20 partially overlap those used here, but sometimes are drawn from other groups' pipelines.  For these reasons, there can be differences in both the periodicity detections (or lack thereof) and the derived periods, which we detail in this section.

\subsection{Discussion of Objects with Periodicities or Quasi-Periodicities Reported Here but Not in R20}

EPIC~210780789 has a clear periodic signature ($P=1.33$d) in our analysis that is not reported in R20. Likewise, EPIC~210777988 (T~Tau) is not reported in R20 but has clear periodic behavior ($P=2.81$d). 
Upon consultation, it appears that these were unintended omissions, rectified in \cite{rebull2020errat}.

We identify two additional stars with (quasi-)periodicities that are not reported in R20. Their identifiers are: EPIC~210865655 and EPIC~211104793.
Both of these sources are bursters with large $Q$ values, indicating that the measured periods are obscured by additional superposed variability, and thus on the edge of detectability.  They are listed as periods here since the light curves satisfy our $Q<0.85$ criterion for repetitive behavior.

\subsection{Discussion of Individual Objects Reported as Periodic in Both Studies}

Among objects for which both we and R20 detect periodicities, there are just two stars with discrepant periods, where the timescale derived here is much longer than the period given by R20. 

The first is EPIC 246925324 or DQ Tau, which has two dominant signals in the light curve. One of these is the $\approx3$ day rotation period of the primary star (quoted in R20; see also the K2-based study of \cite{kospal2018}). The other is a $\approx 15.5$ day repeating signal which is the burst timescale given in Table~2, and equivalent to the orbital period of the binary that induces the bursts \citep{muzerolle2019}. This is a somewhat special system in which each of the two periods is correct, as they measure different physical processes associated with this special system.

The second star, EPIC~247591534, appears quasi-periodic ($Q=0.67$) in our dataset, whereas \citet{rebull2020} report a period of 1.20 days. We do not see any such signal in our periodogram or autocorrelation function.

\subsection{Discussion of Individual Objects with Reported Periodicities Not Recovered in This Study}

Among our sources with $Q>0.85$, for which we do not claim periodicity or quasi-periodicity, there are 15 disk-bearing stars with periods in \citet{rebull2020} that we are either unable to confirm or do not consider as the dominant source of variability.  

\begin{itemize}
\item
EPIC~247885481 is an interesting case in which narrow dips are superimposed on lower amplitude quasi-periodic behavior. We {\em do} detect a signal at the same period as reported by R20 (2.98~d), but the deeper fading events do not phase up on this timescale. We therefore measure a large $Q$ value (0.83) and have labeled this source an aperiodic dipper.
\item
EPIC~246942563 has a potential periodic signal at $P=5.2$~d, as detected by both R20 and us. However, we do not find this period to be significant. The light curve is dominated by erratic short-timescale fading events that do not phase up.
\item
For EPIC~246923113, we detect potential periods around 6, 9 and 15~d, but none of them is significant in the periodogram. R20 report $P=9.0$~d, but given the $Q$ value of 0.87, we consider this light curve to be predominantly stochastic. 
\item
EPIC~248051303 is a borderline case in which we can visualize a repeated brightening pattern in the light curve, but it is noisy and changes amplitude significantly enough that the measured $Q$ value is 0.84. We therefore retain this object as a burster without labeling it quasi-periodic.
\item
EPIC~247584113 or CI Tau is reported by R20 to have a period of 9.0~d.
The $K2$ light curve for this source was also studied in detail by \cite{biddle2021}, who find periods of 6.56~d and 9.06d that they attribute to the rotation period of the star and the pulsed accretion induced by an orbiting planet, respectively, as well as a timescale of 24.4~d.

We see primarily stochastic behavior, with no clearly distinguishable signal on these timescales when the light curve is considered as a whole.  The Q=1.0 and M=0.04 make CI~Tau essentially a highly stochastic source, with sporadic short-timescale ($<$1~day) bursts or flares superimposed. We observe a peak in the autocorrelation function arounnd 19--20~d, but it does not pass our significance threshold in the periodogram and the residuals are large after subtraction of the phased light curve.

\item
EPIC~247047380 is reported by R20 to have a period of 7.4~d, and we see a peak at this location in our periodogram, but it is of low significance. With $Q=0.82$, this light curve is dominated by bursting behavior that changes in amplitude. We therefore do not consider it quasi-periodic.
\item
EPIC~247589612 has a potential periodicity at $P=2.2$~d (as reported by R20) and seen by us in the periodogram), but the light curve is dominated by higher amplitude aperiodic behavior. With $Q=0.94$, we have therefore classified it as an aperiodic dipper, with a secondary designation of quasi-periodic.
\item
EPIC~10690735 is an aperiodic dipper. We see a peak in the autocorrelation function at period 3.68~d, but a $Q=0.9$ it is not significant enough to label quasiperiodic. 
\item 
EPIC~248018652 has a period of 1.16~d reported by \citet{rebull2020}. We do detect a signal near this ($P=1.13$~d) in our periodogram, but the light curve is too noisy for it to be detected as significant. We determine a $Q$ value of 0.96, confirming that noise dominates this light curve.
\item
EPIC~247810751 is reported by \citet{rebull2020} to have a period of 3.28~d; we tentatively detect a signal at roughly double this value ($P=6.56$~d), but it is dominated by higher amplitude aperiodic behavior in the light curve. With $Q=0.92$, we do not list this object as quasi-periodic.
\item
\citet{rebull2020} report a period of 7.8~d for EPIC~248029373, but we do not any periodic signal. This object is therefore classified as stochastic in our list.
\item
EPIC~247604953 has a period of 9.26~d reported by \citet{rebull2020}, but because long-term stochastic behavior dominates this light curve, the phased light curve exhibits a large amplitude spread and the resulting $Q$ is well above our threshold of 0.85.
\item
EPIC~247792225 has a purported period of 7.13~d as listed by \citet{rebull2020}, but we find this signal to be insignificant in the periodogram and dominated by larger amplitude stochastic behavior. The $Q$ value is 0.82.
\item
EPIC~247592463 is a dipper star with a period of 4.33~d listed in R20. Visual inspection of the light curve reveals some dips on this timescale, but their depths vary significantly. The measured $Q$ value is almost 0.9, indicating that the light curve does not phase up cleanly enough to warrant a label of quasi-periodic.
\item
EPIC~210690598 is LkH$\alpha$358, a Class I symmetric source with $\approx3.5$ day rotation period quoted in R20 but our measured timescale is 10.5d at $Q=0.86$.
\end{itemize}

\end{document}